\documentclass[prx,aps,twocolumn,superscriptaddress]{revtex4-2}

\usepackage[pdftex]{graphicx}
\usepackage{amsbsy,amssymb,amsmath,bm,mathtools}
\usepackage{amsfonts}

\usepackage{xspace}
\usepackage{bm}
\usepackage{color}
 
\usepackage{url}
\usepackage[section]{placeins}
\usepackage[colorlinks=true,linkcolor=blue,citecolor=blue]{hyperref}

\usepackage[mathscr]{euscript}

\usepackage{graphicx}
\usepackage{dcolumn}

\begin{document}

\preprint{APS/123-QED}

\title{Machine Learning Force-Field Approach for Itinerant Electron Magnets}

\author{Sheng Zhang}
\affiliation{Department of Physics, University of Virginia, Charlottesville, VA 22904, USA}

\author{Yunhao Fan}
\affiliation{Department of Physics, University of Virginia, Charlottesville, VA 22904, USA}

\author{Kotaro Shimizu}
\affiliation{RIKEN Center for Emergent Matter Science, Saitama 351-0198, Japan} 
\affiliation{Department of Physics, University of Virginia, Charlottesville, VA 22904, USA}

\author{Gia-Wei Chern}
\affiliation{Department of Physics, University of Virginia, Charlottesville, VA 22904, USA}

\begin{abstract}
We review the recent development of machine-learning (ML) force-field frameworks for Landau-Lifshitz-Gilbert (LLG) dynamics simulations of itinerant electron magnets, focusing on the general theory and implementations of symmetry-invariant representations of spin configurations. The crucial properties that such magnetic descriptors must satisfy are differentiability with respect to spin rotations and invariance to both lattice point-group symmetry and internal spin rotation symmetry.  We propose an efficient implementation based on the concept of reference irreducible representations, modified from the group-theoretical power-spectrum and bispectrum methods. The ML framework is demonstrated using the s-d models, which are widely applied in spintronics research. We show that LLG simulations based on local fields predicted by the trained ML models successfully reproduce representative non-collinear spin structures, including 120$^\circ$, tetrahedral, and skyrmion crystal orders of the triangular-lattice s-d models. Large-scale thermal quench simulations enabled by ML models further reveal intriguing freezing dynamics and glassy stripe states consisting of skyrmions and bi-merons. Our work highlights the utility of ML force-field approach to dynamical modeling of complex spin orders in itinerant electron magnets. 
\end{abstract}

\maketitle


\section{Introduction}

\label{sec:intro}

Itinerant frustrated magnets, driven by electron-mediated spin-spin interactions, often give rise to intricate non-collinear and non-coplanar spin textures. Among these, magnetic vortices and skyrmions stand out due to their fundamental significance in magnetism and their promising applications in spintronics~\cite{bogdanov89,rossler06,muhlbaure09,yu10,yu11,seki12,nagaosa13}. These nanometer-scale, localized spin structures are distinguished by nontrivial topological invariants, rendering them exceptionally stable with long lifetimes. In itinerant electron magnets, skyrmions can be dynamically manipulated, generated, and annihilated not only through magnetic fields but also via electrical currents, thanks to the interaction between electron spins and magnetic textures. This capability unlocks exciting possibilities for advanced spintronic technologies. Moreover, complex magnetic textures in itinerant magnets impart a nontrivial Berry phase to electrons as they traverse closed paths around non-collinear or non-coplanar spin configurations~\cite{Batista2016,Hayami2021,Chern2015}. This Berry phase behaves like an effective magnetic field, reshaping the electron band structure and giving rise to distinctive topological responses, such as the anomalous Hall effect.

The geometry of electronic structures, particularly the Fermi surface, plays a pivotal role in shaping the effective spin interactions in itinerant magnets. In the regime of weak electron-spin coupling, these effective spin interactions are closely tied to the particle-hole susceptibility of the electron gas. For a spherical Fermi surface, which is a reasonable approximation for many electron band structures at low filling fractions, this relationship gives rise to the well-known Ruderman-Kittel-Kasuya-Yosida (RKKY) interaction~\cite{Ruderman1954,Kasuya1956,Yosida1957}. Itinerant magnets thus provide a platform for realizing intricate magnetic textures through the engineering of electronic structures. For instance, a skyrmion crystal can be interpreted as a multiple-$\mathbf Q$ magnetic order driven by the Fermi surface nesting mechanism~\cite{Martin2008,Chern2010,Hayami2014,Hayami2018,Wang2020,Hayami2021square}.  By tuning the Fermi surface, for example, through the adjustment of hopping constants or the filling fraction, it is possible to control both the structure and periodicity of a skyrmion crystal.


Modeling the dynamics of complex spin textures in itinerant magnets, however, presents significant computational challenges. The magnetic moments in many metallic skyrmion materials can be reasonably approximated as classical spin vectors, and the dynamics of these classical spins is described by the Landau-Lifshitz-Gilbert (LLG) equation~\cite{LL,Gilbert04}. However, the local effective magnetic fields, analogous to atomic forces in molecular dynamics, arise from exchange interactions with itinerant electrons and must be determined quantum mechanically. Consequently, dynamical simulations require solving an electronic structure problem for the instantaneous spin configuration at each time step. Due to the super-linear computational complexity of most electronic structure methods, repeated calculations of electron structures can become prohibitively expensive in large-scale LLG simulations.

On the other hand, linear-scaling electronic structure methods are feasible if the locality principle, or what Kohn called the ``nearsightedness" principle, is satisfied~\cite{kohn96,prodan05}. This locality principle, stemming from wave-mechanical destructive interference, is widely applicable to electron systems. The nearsightedness of electron systems here does not rely on the existence of well localized Wannier-type wave functions, which only exist in large-gap insulators. Instead, the locality is generally a consequence of wave-mechanical destructive interference of many-particle systems. Importantly, given the locality property, ML force-field models offer an explicit and efficient approach to incorporate the locality principle into the implementation of $\mathcal{O}(N)$ methods. 

Indeed, the locality principle serves as the foundation for recent ML-based force-field methods, enabling large-scale {\em ab~initio} molecular dynamics (MD) simulations with quantum-level accuracy~\cite{behler07,bartok10,li15,shapeev16,behler16,botu17,smith17,zhang18,deringer19,mcgibbon17,suwa19,chmiela17,chmiela18,sauceda20,unke21}. In these methods, atomic forces -- which are central to MD simulations -- are assumed to depend primarily on the local chemical environment. A fixed-size ML model is then trained to encode the highly complex relationship between the force acting on an atom and its local surrounding. A practical implementation of such ML force-field models were demonstrated in the pioneering works of Behler and Parrinello~\cite{behler07} and Bart\'ok {\em et al.}~\cite{bartok10}. Similar ML frameworks have recently been developed to enable large-scale modeling and simulations in several condensed-matter lattice systems~\cite{zhang20,zhang21,zhang22,zhang22b,zhang23,cheng23,cheng23b,Ghosh24,Fan24,Ma19,Liu22,Tian23}. 

Another crucial component, also already emphasized in the original works of  Behler and Parrinello~\cite{behler07}, is the descriptor for proper representation of the chemical neighborhood. Despite powerful approximation capabilities of various learning models, especially deep neural networks, symmetries of the original electron Hamiltonian can only be approximated in the statistical sense. To incorporate symmetry constraints into ML force-field models, the chemical environment should be mapped to a set of feature variables, also known as a descriptor, that are invariant under rotations of the SO(3) symmetry group. The research on atomic descriptors is an active ongoing field, with several different approaches already proposed in the past decades~\cite{behler07,bartok10,li15,behler11,shapeev16,behler16,ghiringhelli15,bartok13,drautz19,himanen20,huo22}.


In this paper, we present a general ML force-field framework for LLG dynamics simulations of itinerant electron magnets, focusing on the theory and implementation of magnetic descriptors. For concreteness, we demonstrate our approach using the s-d model, which is widely used in theoretical modeling of spintronics materials and devices. In contrast to the SO(3) rotation symmetry of molecular systems, the fact that spins in s-d models are arranged on a lattice indicates the relevant symmetry is given by the point-group symmetry of the underlying lattice. On the other hand, in the absence of spin-orbit coupling, the system energy is invariant with respect to simultaneous rotations of all spins. The magnetic descriptor thus needs to account for the lattice point-group symmetry as well as the internal spin rotation symmetry. We will also discuss descriptors for systems with strong spin-orbit coupling, where the rotation in spin space is coupled to the discrete rotation in real-space.

By applying the ML framework to triangular-lattice s-d models, we show that the ML-based LLG simulations accurately reproduce three representative non-collinear spin orders on the triangular lattice: the 120$^\circ$ order with a tripled unit cell, the non-coplanar tetrahedral order with a quadrupled unit cell, and a skyrmion crystal. In particular, large-scale simulations enabled by ML models are crucial for studying the phase ordering dynamics of skyrmion crystals due to their much enlarged magnetic unit cells. Here we consider an s-d model with engineered tight-binding Hamiltonian that exhibits quasi-nesting of its Fermi surface. Although the quasi-nesting mechanism indeed stabilizes a skyrmion crystal, likely as a global minimum, our large-scale thermal quench simulations, however, show that the system  invariably freezes in metastable states consisting of stripe-like structures with skyrmions and bi-merons.




\section{Machine learning force-field model}

\label{sec:ML}

\subsection{Adiabatic dynamics and effective spin Hamiltonians}

We first discuss the governing equations for itinerant electron magnets. For concreteness, we consider the s-d exchange model which describes the interaction between localized magnetic moments ($d$ or $f$ electrons) and itinerant conduction electrons ($s$ electrons) in a material. This model is crucial for understanding magnetic properties in materials where both localized and itinerant electrons contribute to the overall magnetic behavior, such as in transition metals and rare-earth magnetic systems. A general form of the one-band s-d model is described by the following Hamiltonian
\begin{eqnarray}
	\label{eq:H1}
	\hat{\mathscr{H}} = \hat{\mathcal{H}}_e\bigl(\hat{c}^{\,}_{i\alpha}, \hat{c}^\dagger_{i \alpha} \bigr)
	- J \sum_{i, \alpha\beta} \mathbf {S}_i \cdot \hat{c}^\dagger_{i\alpha} \bm\sigma^{\,}_{\alpha\beta} \hat{c}^{\,}_{i \beta},  \quad
\end{eqnarray}
where $\hat{c}_{i \alpha}^{\dagger}(\hat{c}^{\,}_{i \alpha})$ is the creation (annihilation) operators of an electron with spin $\alpha = \uparrow, \downarrow$ at site-$i$. The first  term describes the Hamiltonian for the $s$ electrons, while $J$ in the second term represents the exchange coupling, also known as the Hund's rule coupling, between local magnetic moment $\mathbf S_i$ from the $d$-electrons and spin of $s$-electrons. The electron operators in the second term correspond to the spin operator of the electrons: $\hat{\mathbf s}_i = \frac{\hbar}{2} \bigl(\hat{c}^\dagger_{i, \alpha} \bm\sigma_{\alpha\beta} \hat{c}^{\,}_{i, \beta} \bigr)$. For most spintronics materials, the local magnetic moments are often of large spin length and can be well approximated as classical spins. For convenience, we use a unit system such that the local moments are dimensionless vectors with a unit length $|\mathbf S_i | = 1$.


The dynamics of the classical local spins is governed by the  Landau-Lifshitz-Gilbert equation
\begin{eqnarray}
	\label{eq:LLG}
	\frac{d\mathbf S_i}{dt} = \gamma \mathbf S_i \times ( \mathbf H_i + \bm\eta_i )  
	- \gamma \lambda \mathbf S_i \times \left[ \mathbf S_i \times ( \mathbf H_i + \bm\eta_i ) \right].
\end{eqnarray}
Here  $\gamma$ is the gyromagnetic ratio, $\lambda$ is a dimensionless damping constant, $\mathbf H_i$ is the local effective field exerted by itinerant electrons through the s-d exchange coupling, and $\bm\eta_i = (\eta_i^x, \eta_i^y, \eta_i^z)$ represents a stochastic magnetic field due to thermal fluctuations, which is described by a Gaussian stochastic process with the following statistical properties~\cite{Garcia98}
\begin{eqnarray}
	\langle \bm\eta_i(t) \rangle = 0, \quad
	\langle \eta^m_i(t) \eta^n_j(t')\rangle = \frac{\lambda}{1+\lambda^2} \frac{k_B T}{\gamma}, \quad
\end{eqnarray}
where $T$ is the temperature of the heat bath and $k_B$ is the Boltzmann constant. 

The calculation of the local magnetic field $\mathbf H_i$ that drives the spin dynamics is computationally the most time-consuming part of the LLG simulations. Here we consider the adiabatic dynamics which assumes a much faster electronic relaxation compared with the spin dynamics. This is similar to the Born-Oppenheimer approximation in quantum molecular dynamics (MD)~\cite{Marx09}. Within the adiabatic approximation, the local field is given by the derivative of an effective energy of the system:
\begin{eqnarray}
	\label{eq:local-H}
	\mathbf H_i = -\frac{\partial E}{\partial \mathbf S_i}.
\end{eqnarray}
 Assuming a quasi-equilibrium electron subsystem, the effective energy is given by the expectation value of the Hamiltonian
\begin{eqnarray}
	\label{eq:E_exact}
	E = \bigl\langle \hat{\mathscr{H}}(\{\mathbf S_i\}) \bigr\rangle 
	= {\rm Tr}\bigl(\hat{\rho}_e \hat{\mathscr{H}}  \bigr),
\end{eqnarray}
where $\hat{\rho}_e = \exp(- \hat{\mathscr{H}} / k_B T)/\mathcal{Z}$ is the equilibrium electron density matrix with respect to the instantaneous spin configurations, and $\mathcal{Z}$ is the partition function. For the $s-d$ model in Eq.~(\ref{eq:H1}), the local field can be readily derived as $\mathbf H_i = J \sum_{\alpha\beta} \bm \sigma^{\,}_{\alpha\beta} \rho^{\,}_{i\beta, i \alpha}$, where $\rho^{\,}_{i\alpha, j\beta} \equiv \langle \hat{c}^\dagger_{j\beta} \hat{c}^{\,}_{i\alpha} \rangle$ is the single-electron density matrix.

The total energy $E = E(\{\mathbf S_i \})$ as a function of spins can thus be viewed as an effective classical spin Hamiltonian by integrating out the electron degrees of freedom. In the weak coupling $J \ll W$, where $W$ is the bandwidth of the electronic Hamiltonian $\hat{\mathcal{H}}_e$, a perturbation calculation implies a series expansion in terms of multi-spin interactions
\begin{eqnarray}
	\label{eq:E_spin}
	& & E = \sum_{i,j} J_{ij} \, \mathbf S_i \cdot \mathbf S_j   + \sum_{ijkl} K_{ijkl} (\mathbf S_i \cdot \mathbf S_j) (\mathbf S_k \cdot \mathbf S_l) \nonumber \\
	& & \qquad + \sum_{ijklmn} L_{ijklmn}  (\mathbf S_i \cdot \mathbf S_j) (\mathbf S_k \cdot \mathbf S_l)  (\mathbf S_m \cdot \mathbf S_n)  \\
	& & \qquad + \sum_{ijklmn} M_{ijklmn} (\mathbf S_i \cdot \mathbf S_j \times \mathbf S_k) (\mathbf S_l \cdot \mathbf S_m \times \mathbf S_n) + \cdots \nonumber
\end{eqnarray}
Another four-spin interaction $(\mathbf S_i \times \mathbf S_j) \cdot (\mathbf S_k \times \mathbf S_l)$, which could favor non-coplanar spin configurations, could be absorbed into the $K$-terms above. The various coupling coefficients $J_{ij}$, $K_{ijkl}$, $\cdots$, are further constrained by the lattice symmetry. The expansion at second order gives a bilinear Heisenberg interaction with long-ranged $J_{ij} = \mathcal{J}(r_{ij})$, which is similar to the well-known RKKY interaction~\cite{Ruderman1954,Kasuya1956,Yosida1957}. 
Given the series expansion of Eq.~(\ref{eq:E_spin}), the calculation of the local field is straightforward and can be done very efficiently. For intermediate and large electron-spin coupling, however, higher-order terms in the expansion cannot be neglected. In fact, both biquadratic and six-order terms are shown to play an important role in the stabilization of noncoplanar spin orders, such as the chiral tetrahedral order~\cite{Momoi1997,Akagi2012}. Systematic inclusion of higher order terms, however, is rather tedious, if not impossible.

An alternative approach is to perform LLG simulations by integrating out the electron degrees of freedom on-the-fly, which means an electronic structure problem $\hat{\mathscr{H}}(\mathbf S_i)$ needs to be solved at each time-step of the simulation in order to obtain the density operator $\hat{\varrho}_e(t)$ for the energy and forces calculations; see Eqs.~(\ref{eq:local-H}) and (\ref{eq:E_exact}). This approach is similar to the {\em ab initio} MD method where atomic forces are obtained by solving, e.g. an Kohn-Sham equation at every step of the simulation~\cite{Marx09}. The computational complexity for this electronic structrure calculation of this effective energy depends on the electron Hamiltonian $\hat{\mathcal{H}}_e$ in Eq.~(\ref{eq:H1}). In the absence of electron-electron interaction, $\hat{\mathcal{H}}_e$ corresponds to a tight-binding Hamiltonian and can be solved using the exact diagonalization (ED) with a time complexity $\mathcal{O}(N^3)$, where $N$ is the number of spins, or linear-scaling techniques such as kernel polynomial methods (KPM)~\cite{Weisse06,Barros13,Wang18}.
For itinerant magnets with interacting electrons, such as Hubbard repulsion, more sophisticated many-body methods are required, often with a super-linear computational complexity. As a result, large-scale LLG simulations  would be prohibitively expensive with repeated many-body calculations.

\begin{figure*}[t]
\centering
\includegraphics[width=2\columnwidth]{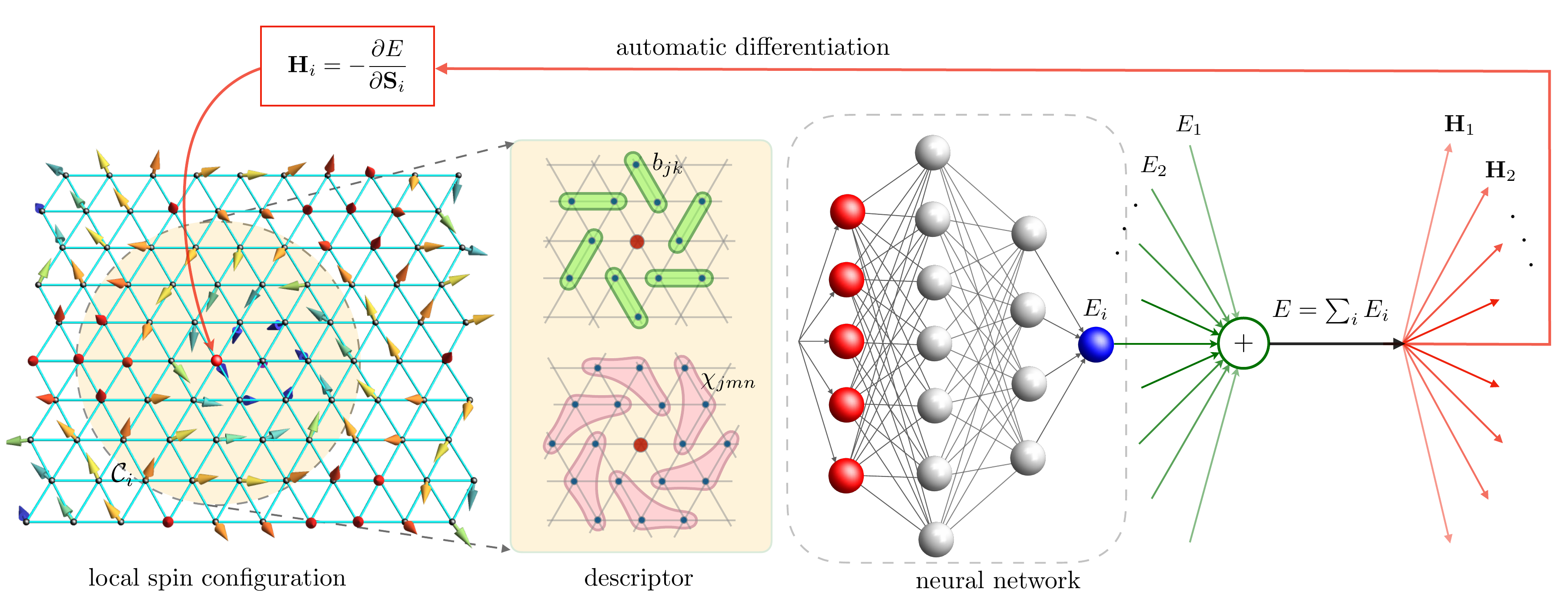}
\caption{Schematic diagram of ML force-field model for itinerant electron magnets. A descriptor transforms the neighborhood spin configuration $\mathcal{C}_i$ to effective coordinates $\{G_\ell \}$ which are then fed into a neural network (NN). The output node of the NN correspond to the local energy $\epsilon_i = \varepsilon(\mathcal{C}_i)$ associated with spin $\mathbf S_i$. The corresponding total potential energies $E$is obtained from summation of these local energies. Automatic differentiation is employed to compute the derivatives $\partial E / \partial \mathbf S_i$ from which the local exchange fields $\mathbf H_i$ are obtained. }
    \label{fig:ml-scheme}
\end{figure*}

\subsection{Behler-Parrinello ML architecture}

As discussed above, linear-scaling electronic structure methods are possible mainly because of the locality nature or ``nearsightedness'' principle of many-electron systems~\cite{kohn96,prodan05}. For example, locality principle is implicitly utilized in the linear-scaling KPM for approximating spectral properties of large bilinear fermionic Hamiltonians~\cite{Weisse06,Barros13,Wang18}. By expanding, e.g. the electron density of states in terms of Chebyshev polynomials, the expansion coefficients can be recursively obtained through repeated sparse matrix-vector multiplications with a $\mathcal{O}(N)$ complexity. The sparsity of the Hamiltonian matrix is a direct consequence of the locality property.  However, KPM can only be used for systems without electron-electron interactions, as described by bilinear fermionic Hamiltonians. For most electronic structure methods for correlated electrons, it is unclear how to integrate the locality principle to improve the computational efficiency. 

On the other hand, ML force field framework offers a systematic approach to achieve linear scalability by explicitly taking advantage of the locality property. An explicit implementation for {\em ab initio} MD simulations was demonstrated in the pioneer work of Behler and Parrinello (BP)~\cite{behler07}. Here we generalize the BP scheme to the case of itinerant electron magnets. The locality principle indicates that the effective field $\mathbf H_i$ associated with a local spin $\mathbf S_i$ only depends on the immediate surrounding of the $i$-th site. For convenience, we introduce the notation $\mathcal{C}_i$ to denote spin configuration in a finite neighborhood of site-$i$,
\begin{eqnarray}
	\mathcal{C}_i = \{ \mathbf S_j \,  \bigl| \, |\mathbf r_j - \mathbf r_i| < r_c \},
\end{eqnarray}
where the cutoff radius $r_c$ depends on the locality. The universal mapping  (in the sense of the given s-d model) from $\mathcal{C}_i$ to the effective field $\mathbf H_i$ can then be approximated by a ML model. Yet, instead of directly predicting the forces, the BP scheme focuses on the energy prediction. As discussed in the next section, this approach could better facilitate the incorporate of symmetry constraint. Another advantage of this scheme is the explicit implementation of a conservative force as implied in Eq.~(\ref{eq:local-H}). 

A schematic diagram of the BP-type ML force-field model is shown in Fig.~\ref{fig:ml-scheme} for the spin dynamics. First, the effective energy $E$, formally obtained by integrating out the electrons in Eq.~(\ref{eq:E_exact}), is decomposed into local energies $\epsilon_i$ associated with individual spins:
\begin{eqnarray}
	\label{eq:E_ML}
	E = \sum_i \epsilon_i = \sum_i \varepsilon(\mathcal{C}_i).
\end{eqnarray}
As shown in the second equality above, we apply the principle of locality, assuming that the local energy $\epsilon_i$ depends solely on its immediate magnetic environment $\mathcal{C}_i$, through a universal function $\varepsilon(\cdot)$. The intricate relationship between local energy and the spin environment $\mathcal{C}_i$ is approximated using a machine learning model.  
For ML force-field models for {\em ab initio} MD simulations, various approaches have been introduced to describe the potential-energy surface as a function of the local chemical or atomic environment, with prominent examples including the Gaussian Approximation Potential~\cite{bartok10} and Neural Network Potentials~\cite{behler07,zhang18}. In this work, we employ feedforward neural networks (NNs) as the learning model, which are capable of accurately representing highly complex functions with arbitrary precision thanks to the universal approximation theorem~\cite{Cybenko89,Hornik89}.


The total energy $E$, given by the sum of local energies, is obtained by applying the same NN model to each individual spin in the lattice. This process is analogous to assembling identical ML models into a ``super neural network" that takes the entire spin configuration $\{ \mathbf S_i \}$ as input and outputs the total energy $E$ at its final node, as shown in Fig.~\ref{fig:ml-scheme}. The effective local field $\mathbf H_i$, defined as the derivative of the total energy as shown in Eq.~(\ref{eq:local-H}), can be efficiently computed using automatic differentiation techniques~\cite{Paszke17,Baydin18}. In practice, however, due to the locality of the effective field, the calculation of $\mathbf H_i$ only requires contributions from local energies $\epsilon_j$ within a similar finite neighborhood around site-$i$

The feedforward neural network, illustrated in Fig.~\ref{fig:ml-scheme}, is composed of multiple artificial neurons, or nodes, organized into distinct layers: an input layer, one or more hidden layers, and an output layer.  As data propagate sequentially from the input layer to the output layer through the hidden layers, each neuron in a given layer receives a set of inputs $\{ x_m \}$ from the neurons in the preceding layer.  These inputs are processed through a weighted sum, followed by the addition of a bias term, and then passed through a nonlinear activation function. Mathematically, this transformation can be expressed as: $y = f\left( \sum_m w_m x_m + b \right)$, where $w_m$ represents the learnable weights associated with each input connection, $b$ denotes the bias term, and $f(\cdot)$ is the activation function. The activation function introduces nonlinearity into the model, enabling the network to approximate complex relationships between the input and output variables effectively. Common choices for activation functions include sigmoid, tanh, and rectified linear unit (ReLU). During training, the NN adjusts the trainable parameters $\bm \theta = \{ w_m, b \}$ iteratively using optimization algorithms, such as stochastic gradient descent, to minimize a loss function that quantifies the difference between prediction and ground truth.

\section{Magnetic descriptors}

\label{sec:descriptor}

Despite the universal approximation capability of NN, symmetry properties of the original electron Hamiltonian can only be captured approximately. More specifically, symmetry constraints indicate that the NN model fed by two symmetry-related neighborhood configurations $\mathcal{C}_i$ and $\mathcal{C}'_i$, should produce exactly the same local energy $\epsilon_i$ at the output node. As discussed above, symmetry properties can be exactly incorporated into the ML model through a proper representation of the neighborhood. Essentially, instead of directly feeding the spin configuration $\mathcal{C}_i$ into the NN, the input to the NN is a set of generalized coordinates $\{G_1, G_2, \cdots \}$ built from the neighborhood spins:
\begin{eqnarray}
	\label{eq:E_G}
	E = \sum_i \varepsilon_{\bm \theta}\bigl( G_\ell(\mathcal{C}_i) \bigr),
\end{eqnarray}
where the subscript $\bm \theta = \{ w, b \}$ denotes trainable parameters of the NN representation of the energy function. By constructing these feature variables $\{ G_\ell \}$ in such a way that they are invariant under symmetry transformations, Eq.~(\ref{eq:E_G}) indicates that both local energies and system energy are also invariants of the symmetry group. 

The significance of descriptors in developing ML force-field models for {\em ab initio} MD was also highlighted in the pioneering works of Behler and Parrinello~\cite{behler07}. They introduced a set of feature variables known as atom-centered symmetry functions (ACSFs) to encode local atomic environments while explicitly preserving rotational and reflection symmetries in the atomic energy function~\cite{behler07,behler16}. ACSFs are constructed from relative distances and angles between atomic position vectors, inherently ensuring invariance under the SO(3) rotational symmetry group~\cite{behler16}.  Numerous other atomic descriptors have since been proposed and successfully implemented~\cite{bartok10,li15,shapeev16,behler11,ghiringhelli15,bartok13,drautz19,himanen20,huo22}. Notably, a systematic approach to atomic descriptors has been proposed based on the group-theoretical bispectrum method~\cite{bartok10,bartok13}.

Regarding the symmetry group characterizing the neighborhood, there are two main differences between the molecular systems and the s-d spin models. First, contrary to the continuous rotations around a center atom, the SO(3) symmetry is reduced to a discrete point-group symmetry associated with a lattice site. Second, while MD systems are characterized by a permutation symmetry of atomic species, the s-d Hamiltonian is invariant under simultaneous rotations of all spins, i.e. a global SO(3) spin rotation symmetry. As a result, despite extensive works on atomic descriptors, most cannot be directly applied to itinerant magnetic systems.  Feature variables of the magnetic environment $\mathcal{C}_i$ need to be invariants of a symmetry group that is the direct product of global SO(3) of spin rotation and discrete point group of site-symmetry.

\subsection{Global SO(3) spin rotation symmetry}

We first consider feature variables that are invariant under the global spin rotation symmetry. Here the rotation symmetry refers to the global rotation of local magnetic moments $\mathbf S_i \to \mathcal{R}\cdot \mathbf S_i$ (treated as classical vectors), and a simultaneous unitary transformation of the electron spinor $\hat{c}_{i\alpha} \to \hat{U}_{\alpha\beta} \hat{c}_{i\beta}$, where $\mathcal{R}$ is an orthogonal $3\times 3$ matrix and $\hat{U} = \hat{U}(\mathcal{R})$ is the corresponding $2\times 2$ unitary rotation operator. The local energy function $\varepsilon(\mathcal{C}_i)$, which can be viewed as an effective energy after integrating out electrons, should retain the global SO(3) rotation symmetry of classical spins $\mathbf S_i$. This rotation symmetry can be manifestly maintained if the energy function only depends on the bond variables $b_{jk}$ and scalar chirality $\chi_{jmn}$ of spins within the neighborhood; they are defined as
\begin{eqnarray}
	\label{eq:building-blocks}
	b_{jk} = \mathbf S_j \cdot \mathbf S_k, \qquad \chi_{jmn} = \mathbf S_j \cdot \mathbf S_m \times \mathbf S_n.
\end{eqnarray}
These two building blocks also correspond to two-spin and three-spin correlations. Higher-order multi-spin correlations that are invariant under rotations can be expressed in terms of product of $b_{jk}$ and $\chi_{jmn}$. The dependence of local energy on higher-order spin-correlations is realized through the nonlinearity built in the feed-forward neural network. For $M$ spins in the neighborhood $\mathcal{C}_i$, the number of bond and chirality variables scale as $\mathcal{O}(M^2)$ and $\mathcal{O}(M^3)$, respectively. In order to keep a manageable number of these variables, we introduce another cutoff distance $l_c$ such that bond or chirality variables including two off-center spins at site-$m$ and $n$ are considered only when $|\mathbf r_m - \mathbf r_n| \le l_c$. Given the building blocks of Eq.~(\ref{eq:building-blocks}), next we discuss approaches to incorporate the discrete lattice symmetry.

\subsection{Site-centered symmetry function}

\label{sec:scsf}

As a warm up, we discuss an intuitive descriptor for the point-group symmetry, which is similar to the well-known ACSF atomic descriptor~\cite{behler07,behler16}. Consider a feature variable built from a linear combination of either the bond or chirality variables, i.e. $G = \sum_{jk} \mathtt{c}_{jk} b_{jk}$ or $G' = \sum_{jmn} \mathtt{c}_{jmn} \chi_{jmn}$. The coefficients $\mathtt{c}$ in these expressions are not all independent because of the symmetry constraints. Instead of explicitly enumerating the independent coefficients for all possible bond pairs or chirality triplets, an alternative approach is to specify these coefficients based on a few geometrical parameters in a symmetric way. To this end, we first introduce a two-point function
\begin{eqnarray}
	\label{eq:F2-envelop}
	F^{(2)}_{ij}(r, w) = e^{- ( r_{ij} - r )^2/w^2 } \, f_c(r_{ij}).
\end{eqnarray}
Here $r_{ij} = \left| \mathbf r_i - \mathbf r_j \right|$ is the distance between sites $i$ and~$j$, the two parameters $d$ and $w$ specify the center and width, respectively, of the Gaussian function, and $f_c(r) = \frac{1}{2} \bigl[ \cos(\frac{\pi r}{r_c}) + 1 \bigr]$ for $r \le r_c$ and zero otherwise. Here instead of a hard cutoff in the definition of the neighborhood $\mathcal{C}_i$, the function $f_c(r)$ is introduced for a soft constraint of the locality. A symmetry function involves bond variable between the center spin $\mathbf S_i$ and neighboring spins $\mathbf S_j$ is defined as
\begin{eqnarray}
	G_2(r, w) = \sum_j F^{(2)}_{ij}(r, w) (\mathbf S_i \cdot \mathbf S_j).
\end{eqnarray}
This feature variable thus characterizes correlation between the center spin and those within a ring of width $w$ at a distance $d$ from the center site-$i$. By covering the neighborhood $\mathcal{C}_i$ with a collection of such rings with radius $d_m$ and width $w_m$, the set of feature variables $\{ G_2(d_m, w_m) \}$ thus provides a symmetry-invariant representation of the spin correlations within the neighborhood. 
 
In order to characterize off-center spin correlations, namely bond variables $b_{jk}$ that do not involve the center spin $\mathbf S_i$, we first introduce 3-body symmetry functions 
\begin{eqnarray}
	\label{eq:F3-envelop}
	& & F^{(3)}_{ijk}(r, w, d, \delta) =  e^{-\frac{(r_{ij} - r)^2 + (r_{ik} - r)^2}{w^2}}  \\
	& & \qquad \qquad \qquad \times \, e^{-\frac{(r_{jk} - d)^2}{\delta ^2}}\, f_c(r_{ij}) f_c(r_{ik}) . \nonumber
\end{eqnarray}
The various Gaussian functions are used to introduce soft distance-constraints between neighboring pairs. Another symmetry function involving off-center spin pairs is defined as
\begin{eqnarray}
	G_2'(r, w, d, \delta) = \sum_{j,k \neq i} F^{(3)}_{ijk}(r, w, d, \delta) (\mathbf S_j \cdot \mathbf S_k). \quad
\end{eqnarray}
This symmetry-invariant variable describes the average correlation between two spins separated by a distance $\sim d \pm \delta $ that lie within a ring of width $w$ and radius $r$. The three-body function can also be used to construct feature variables involving the chirality of three spins
\begin{eqnarray}
	G_3(r, w, d, \delta) = \sum_{j,k \neq i} F^{(3)}_{ijk}(r, w, d, \delta) (\mathbf S_i \cdot \mathbf S_j \times \mathbf S_k). \qquad
\end{eqnarray}
Since the two- and three-body functions are defined entirely in terms of distances, which are manifestly invariant under rotations/inversions of lattice point groups, the resultant feature variables $G$'s are also invariants with respect to the symmetry group. A symmetry-invariant representation of the magnetic environment $\mathcal{C}_i$ can be obtained by using these symmetry functions with proper parameters to cover the neighborhood.

The above construction of the expansion coefficients is similar, at least in spirit, to the ACSF atomic descriptor. The parametrization offers a systematic way to cover the neighborhood. The formulation also provides a natural way to include the spin degrees of freedom. However, while the lattice setting makes the incorporation of symmetry relatively easier compared withe the continuous rotation symmetry of MD systems, as will be discussed next, the site-centered symmetry functions correspond to only a special subset of possible symmetry-invariant feature variables. Before concluding this section, we note in passing that a similar magnetic descriptor, combining symmetry functions with bond or chirality spin variables, has been combined with ML force-field models for itinerant electron spin-glasses~\cite{shi23}.

\subsection{Power spectrum and bispectrum descriptors}

\begin{figure}
\centering
\includegraphics[width=0.99\columnwidth]{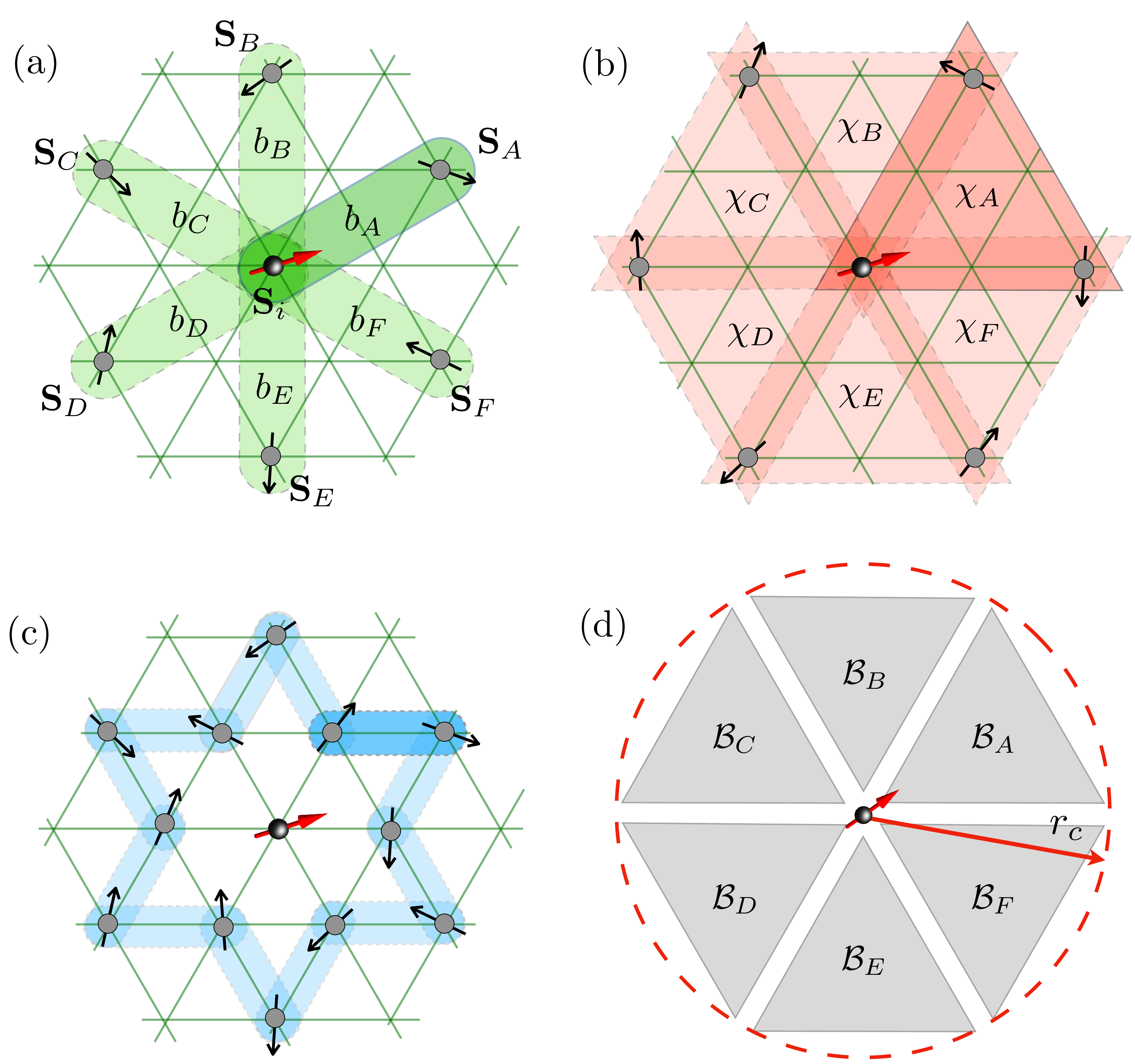}
\caption{Symmetry-invariant descriptor for the neighborhood spin configuration. (a) The six bond variables $b_k = \mathbf S_0 \cdot \mathbf S_k$, constructed from the inner product of the six third-nearest neighboring spins $\mathbf S_k$ and the center spin $\mathbf S_0$, form the basis of a four-dimensional reducible representation of the D$_6$ point group. (b) Similarly, the six scalar chirality variables $\chi_1, \chi_2, \cdots ,\chi_6$, each indicated by a triangle, also form the basis of a reducible 6-dim representation of $D_6$. The scalar chirality of a triangle $(ijk)$ is  $\chi = \mathbf S_i \cdot \mathbf S_j \times \mathbf  S_k$. (c) An example of neighboring off-site bond variables forming a 12-dim reducible representation of the $D_6$ group. (d) Schematic diagram showing six symmetry-related blocks of neighboring sites within the cutoff radius $r_c$ of a center spin.}
    \label{fig:descriptor}
\end{figure}

A more systematic approach to obtain invariant variables is based on group-theoretical bispectrum method~\cite{kondor07,bartok13}. Here we outline how this works for the lattice point-group symmetry of magnetic systems. First we note that the set of all bond and chirality variables in the neighborhood $\mathcal{C}_i$ form the basis of a large dimensional irreducible representation of the point group associated with the site symmetry. This basis set can be decomposed to fundamental irreducible representations (IR’s), which are building blocks for forming invariants, of the point group. Although there is well-established procedure for obtaining these IRs, the decomposition can be highly simplified as the original representation matrix is automatically block-diagonalized. This is because the distance between a spin-pair (bond) or spin-triplet (scalar chirality) from the center site is preserved by operations of the discrete point group. Each block then corresponds to $b_{jk}$ or $\chi_{jkl}$ of a fixed distance; see Fig.~\ref{fig:descriptor}(a)--(c). 

Specifically for the case of triangular lattice considered in this work, the point group characterizing the site symmetry is $D_6$. The dimension of each block of the block-diagonalized representation matrix is either 6 or 12. Take for example the six third-nearest neighboring spins $\mathbf S_A, \mathbf S_B, \cdots, \mathbf S_F$ in Fig.~\ref{fig:descriptor}(a). The resultant six bond variables $b_\mu = \mathbf S_i \cdot \mathbf S_\mu$ ($\mu = A, B, \cdots, F)$ then form the basis of a dimension-6 representation of the $D_6$ group, and can be decomposed into $6 = A_1 \oplus B_2 \oplus E_1 \oplus E_2$, with the following basis~\cite{hamermesh62}:
\begin{equation} \label{eq1}
\begin{split}
f^{A_1} & = b_{A} + b_{B} + b_{C} + b_{D} + b_E + b_F \\
f^{B_2} & = b_{A} - b_{B} + b_{C} - b_{D} + b_E - b_F \\
f^{E_1}_x & = \frac{\sqrt{3}}{2} ( b_{B} + b_{C} - b_E - b_F) \\
f^{E_1}_y & = \frac{1}{2} ( 2b_A + b_B - b_C - 2 b_D  -b_E + b_F) \\
f^{E_2}_x & = \frac{1}{2} ( 2b_A - b_B - b_C + 2 b_D - b_E - b_F) \\
f^{E_2}_y & = \frac{\sqrt{3}}{2} (-b_B + b_C - b_E + b_F)
\end{split}
\end{equation} 
Another example is the 6-dimensional reducible representation based on the six scalar chirality variables $\chi_A, \chi_B$, $\cdots$, $\chi_F$ that depend on the center spin $\mathbf S_i$ and 6 neighboring spins as shown in Fig.~\ref{fig:descriptor}(b). In this case, since the scalar chirality variables change sign under an odd permutation of the three spins, their decomposition is given by $6 = A_2 \oplus B_1 \oplus E_1 \oplus E_2$ with the basis 
\begin{equation} \label{eq2}
\begin{split}
f^{A_2} & = \chi_{A} + \chi_{B} + \chi_{C} + \chi_{D} + \chi_E + \chi_F \\
f^{B_1} & = \chi_{A} - \chi_{B} + \chi_{C} - \chi_{D} + \chi_E - \chi_F \\
f^{E_1}_x & = \frac{\sqrt{3}}{2} ( \chi_{B} + \chi_{C} - \chi_E - \chi_F) \\
f^{E_1}_y & = \frac{1}{2} ( 2 \chi_A + \chi_B - \chi_C - 2 \chi_D  - \chi_E + \chi_F) \\
f^{E_2}_x & = \frac{1}{2} ( 2 \chi_A - \chi_B - \chi_C + 2 \chi_D - \chi_E - \chi_F) \\
f^{E_2}_y & = \frac{\sqrt{3}}{2} (- \chi_B + \chi_C - \chi_E + \chi_F)
\end{split}
\end{equation}
Finally, we give an example of a subspace of 12 dimensions, shown in Fig.~\ref{fig:descriptor}(c). In this case, the basis of this reducible representation is given by the off-site bond variables (which means the center spin is not involved) from 12 neighboring spins. The decomposition of this 12-dimensional representation is $12 = A_1 \oplus A_2 \oplus B_1 \oplus B_2 \oplus 2\, E_1 \oplus 2\, E_2$.

By applying the decomposition to all blocks, we arrange the resultant IR coefficients into a vector 
\begin{eqnarray}
	\bm f^{(\Gamma, r)} = \Bigl(f^{(\Gamma, r)}_1, f^{(\Gamma, r)}_2, \cdots, f^{(\Gamma, r)}_{n_\Gamma} \Bigr), 
\end{eqnarray}
where $\Gamma$ denotes the type of IR, $n_\Gamma$ its dimension, and $r$ enumerates the multiplicity of the IR. These IR coefficients are the point-group analogs of Fourier components of the O(2) group, or the spherical components $Y_{\ell, m}$ of the SO(3) rotation group. Importantly, the transformation of each vector under symmetry operations is completely determined by the known matrix representation $\bm D^{(\Gamma)}$ associated with the IR: $\bm f^{(\Gamma, r)} \to \bm D^{(\Gamma)} \cdot \bm f^{(\Gamma, r)}$. This property thus allows us to systematically construct feature variables that are invariant with respect to the point-group symmetry.

For example, similar to the ordinary Fourier analysis, the so-called power spectrum of IR is invariant under the symmetry operations:
\begin{eqnarray}
	p^{(\Gamma, r)} = \bm f^{(\Gamma, r) \, \dagger} \cdot \bm f^{(\Gamma, r)} = \sum_{\mu = 1}^{n_\Gamma} \bigl| f^{(\Gamma, r)}_{\mu} \bigr|^2.
\end{eqnarray}
However, a descriptor based only on power spectrum is obviously incomplete in the sense that the phase of 1D IR or directional information of higher-dimensional IR is lost. Moreover, as the power spectrum is invariant even under transformations of a single IR (with all others remain fixed), the corresponding symmetry, which is given by the direct product of symmetry operations of each IR, is artificially enlarged compared with the actual point group symmetry. What is missing here is the relative ``phases" between different IRs, which are also invariant of the point group. Take two doublet IRs $\bm f^{(E, 1)}$ and $\bm f^{(E, 2)}$ for example, under symmetry operation, they behave as 2D vectors in a plane. Consequently, not only the length of individual vectors $|\bm f^{(E, r)}|$ is unchanged under rotations, so does the relative angle between them $\cos\theta_{12} \propto \bm f^{(E, 1)} \cdot \bm f^{(E, 2)}$. 
In order to include this additional information, a complete description can be achieved using the so-called bispectrum coefficients~\cite{kondor07,bartok13}
\begin{eqnarray}
	\label{eq:bispectrum}
	b^{(\Gamma, \Gamma_1, \Gamma_2)}_{r, r_1, r_2} = \sum_{\kappa, \mu, \nu} C^{(\Gamma, \Gamma_1, \Gamma_2)}_{\kappa, \mu\nu} f^{(\Gamma, r) \, *}_\kappa f^{(\Gamma_1, r_1)}_\mu f^{(\Gamma_2, r_2)}_\nu, \quad
\end{eqnarray}
where $C^{(\Gamma, \Gamma_1, \Gamma_2)}_{\kappa, \mu\nu}$ are the Clebsch-Gordan coefficients of the point group. The bispectrum coefficients are triple products of coefficients of the same or different IRs, similar to the scalar product $\mathbf a \cdot \mathbf b \times \mathbf c$ of three vectors.   When one of the three IRs, say $\Gamma$, is the trivial identity IR, the resultant bispectrum coefficient is reduced to the inner product of two IRs of the same type. This also means that the power spectrum is a subset of the bispectrum coefficients. 

An atomic descriptor based on the bispectrum coefficients has been combined with Gaussian kernel potential learning model for {\em ab initio} MD simulations~\cite{bartok10,bartok13}. A direction approach to obtain the IR coefficients is to first partition the atomic neighborhood into many partially overlapping shells and apply the decomposition of each shell into the IR of SO(3) group. Instead of dealing with the tedious bookkeeping of multiple shells, an alternative approach is to map the atomic neighborhood onto the surface of a 4-dimensional sphere, which is then decomposed into IR of the SO(4) group~\cite{bartok10,bartok13}. 
Back to the lattice systems, the dimensions $D_\Gamma$ of most IRs are rather small for most point groups, which means there is a large multiplicity (indexed by $r$) for each IR. This in turn results in a large number of possible bispectrum coefficients. Moreover, the set of all bispectrum coefficients listed in Eq.~(\ref{eq:bispectrum}) is over-complete. Consequently, further simplification is required for practical implementations.

\subsection{Reference irreducible representations}

In order to keep the number of feature variables manageable, here we apply the idea of reference IR~\cite{Ma19,zhang22} to represent the magnetic environment. Essentially, the reference IRs are to provide the phase information that is missing in the power spectrum descriptor. The relative phase between two IR coefficients can be inferred from their phases with respect to the reference IR. In order to accommodate IRs of different transformation properties, the reference needs to contain all possible IR types of the symmetry group. Moreover, another desirable property of the reference IRs is the insensitivity to small changes of the magnetic environment. To this end, we compute the reference basis $\bm f^{\Gamma}_{\rm ref}$ for each IR of the point group by averaging large blocks of bond and chirality variables. 

An example of such blocks is shown in Fig.~\ref{fig:descriptor}(d). First, the neighborhood $\mathcal{C}_i$ of a given center site-$i$ is partitioned into six blocks, labeled by $\mathcal{B}_K$, where $K = A, B, \cdots, F$, which are related by symmetries of the $D_6$ group. These six blocks are  indicated schematically by the triangular-shaped regions in the figure. In practical implementations, they can partially overlap with each other. Take the case of bond variables for instance, one can associated with block with an averaged bond variables as $\overline{b}_{K} = \frac{1}{N_b} \sum_{(ij) \in \mathcal{B}_K} b_{ij}$, where $N_b$ is the total number of pairs $(ij)$ in a block. A set of reference IR bases can then be obtained via the decomposition of these six block-averaged bond variables $\overline{b}_K$ according to Eqs.~(\ref{eq1}) or (\ref{eq2}), depending whether the reference is for bond or chirality variables.  

Next we discuss the explicit construction of invariant feature variables for each IR type, again using $D_6$ as an example. A vector of feature variables $\bm G^{(\Gamma, r)}$ of the same dimension $n_\Gamma$ is derived from the corresponding IR coefficient $\bm f^{(\Gamma, r)}$ and the reference. First, since the $A_1$ IR $f^{(A_1, r)}$, which is symmetric sum of the bond/chirality variables, is automatically invariant under symmetry operations, the corresponding feature variable is simply the coefficient itself: 
\begin{eqnarray}
	\label{eq:G_A1}
	G^{(A_1, r)} = f^{(A_1, r)}. 
\end{eqnarray}
In fact, the site-centered symmetry functions discussed in Sec.~\ref{sec:scsf} are special sets of the $A_1$ IRs arranged in a geometrically intuitive way.  

For other 1D IRs, $\Gamma = A_2$, $B_1$, and $B_2$, a phase $\eta = \pm 1$ is defined through the reference IR of the same 1D type:
\begin{eqnarray}
	\eta^{(\Gamma, r)} = f^{(\Gamma, r)} f^{\Gamma}_{\rm ref} / \bigl| f^{(\Gamma, r)} \bigr| \bigl|  f^{\Gamma}_{\rm ref} \bigr|. 
\end{eqnarray}
As the two $f$ coefficients transform in exactly the same way, this phase variable is invariant under symmetry transformations. The corresponding feature variables are then given by
\begin{eqnarray}
	\label{eq:G_1D_IR}
	G^{(\Gamma, r)} = \sqrt{ p^{(\Gamma, r)} } \eta^{(\Gamma, r)}, \qquad (\Gamma = A_2, B_1, B_2), \quad
\end{eqnarray}
where $p^{(\Gamma, r)} = \bigl| f^{(\Gamma, r)} \bigr|^2$ is the power spectrum coefficient. As both $p$ and $\eta$ are invariants, the feature variable $G^{(\Gamma, r)}$ itself is manifestly invariant under symmetry operations. 

Next we consider the doublet IRs. Simple counting indicates that two feature variables $\bm G^{(E, r)} = \bigl(G^{(E, r)}_1, G^{(R, r)}_2 \bigr)$ are required in order to properly reconstruct a doublet IR coefficient $\bm f^{(E, r)}$. There are two approaches to obtain these invariant feature variables. The first approach is similar to the case of nontrivial 1D IRs discussed above. One of the feature variable is provided by the power spectrum coefficient $p^{(\Gamma, r)}$, while the other is given by the relative ``angle" with the reference doublet:
\begin{eqnarray}
	\cos\theta^{(E, r)} = \bigl( \bm f^{(E, r)} \cdot \bm f^{E}_{\rm ref} \bigr) / \bigl|\bm f^{(E, r)} \bigr| \bigl| \bm f^{E}_{\rm ref} \bigr|.
\end{eqnarray}
This relative angle can then be combined with the power spectrum coefficient to obtain a complex feature variable $G^{(E, r)} = \sqrt{ p^{(E, r)} } \exp( i \theta^{(E, r)} )$, or equivalently a two-component vector:
\begin{eqnarray}
	\bm G^{(E, r)} = \sqrt{ p^{(E, r)} } \Bigl( \cos\theta^{(E, r)}, \sin\theta^{(E, r)} \Bigr)
\end{eqnarray} 
It is worth noting that the two independent feature variables here are the power spectrum (amplitude) and relative phase (angle). However, this approach with only one invariant angle built from the reference cannot be easily generalized to higher-dimensional IRs, such as triplet, without losing information. 

A more elegant and general method is based on the bispectrum method. Consider two doublet IRs: $\bm f^a$ and $\bm f^b$, their direct product can be decomposed as $E \otimes E = A_1 \oplus A_2 \oplus B_1 \oplus B_2$, which means, in addition to the inner product $\bm f^a \cdot \bm f^b$, there are three other scalars that can be built from them:
\begin{eqnarray*}
	-i \bm f^a   \cdot   \bm \sigma_2  \cdot  \bm f^b, \qquad \bm f^a  \cdot  \bm \sigma_3  \cdot  \bm f^b, 
	\qquad \bm f^a  \cdot  \bm \sigma_1  \cdot \bm f^b.
\end{eqnarray*}
where $(\bm \sigma_1, \bm\sigma_2, \bm\sigma_3)$ is a vector of the $2\times 2$ Pauli matrices. Under symmetry transformations, these three scalars behave as an $A_2$, $B_1$, and $B_2$ IR, respectively. An truly invariant scalar can then be obtained by combining these with the corresponding 1D IR. For example, a two-component invariant feature variables for a given double IR can be defined as
\begin{eqnarray}
	\label{eq:G_E}
	\bm G^{(E, r)} = \Bigl( \bm f^{(E, r)} \cdot \bm f^{E}_{\rm ref} \, , \,\,
	-i f^{A_2}_{\rm ref} \,  \bm f^{(E, r)} \cdot \bm\sigma_2 \cdot \bm f^{E}_{\rm ref} \Bigr). \qquad
\end{eqnarray}
The second component is essentially a bispectrum coefficient Eq.~(\ref{eq:bispectrum}) obtained with one IR from the target doublet $\bm f^{(E, r)}$ and two from the reference $\bm f^E_{\rm ref}$.  
Similar approach can be used to build invariant 3-component feature variables for triplet IR. And finally, the relative phases between different irreps are encoded in the bispectrum coefficient of the reference irreps, 
\begin{eqnarray}
	b^{(\Gamma, \Gamma_1, \Gamma_2)}_{\rm ref} = \sum_{\kappa, \mu, \nu} C^{(\Gamma, \Gamma_1, \Gamma_2)}_{\kappa, \mu\nu} g^{(\Gamma) \, *}_\kappa g^{(\Gamma_1)}_\mu g^{(\Gamma_2)}_\nu.
\end{eqnarray} 
The various steps of the magnetic descriptor for representing the neighborhood $\mathcal{C}_i$ is summarized here
\begin{eqnarray}
	\{\mathbf S_j\} \, \to \, \{b_{jk}, \chi_{jmn} \} \, \to \, \{\bm f^{(\Gamma, r)}\} \, \to \, \{ \bm G^{(\Gamma, r)}, b^{(\Gamma, \Gamma_1, \Gamma_2)}_{\rm ref}  \}. \nonumber
\end{eqnarray} 
The final feature variables given by Eqs.~(\ref{eq:G_A1}), (\ref{eq:G_1D_IR}), and (\ref{eq:G_E}) for the different IRs are now invariant under both the SO(3) spin rotation and the lattice symmetry operations, and are the input to the NN model as indicated in Eq.~(\ref{eq:E_G}); also see Fig.~\ref{fig:ml-scheme}.

\section{Benchmarks}

In this section we develop ML force-field models to study three representative magnetic orders in the triangular-lattice s-d model. A magnetic descriptor based on the reference IR method is used to incorporate the symmetry of the s-d model into the ML model. As a proof of principle, we consider an s-d model of free electrons, which means the first term in Eq.~(\ref{eq:H1}) is given by a tight-binding Hamiltonian 
\begin{eqnarray}
	\hat{\mathcal{H}}_e = \sum_{ij, \alpha} t_{ij} \hat{c}^\dagger_{i\alpha} \hat{c}^{\,}_{j \alpha}. 
\end{eqnarray}
For a given configuration $\{ \mathbf S_i \}$ of classical spins, the s-d model with tight-binding Hamiltonian for $\hat{\mathcal{H}}_e$ can be solved using either ED or KPM to obtain dataset for training the ML models. In particular, the nearest-neighbor s-d model has been extensively studied theoretically~\cite{Martin2008,Akagi10,Kato2010,Azhar17,Chern2012,Nandkishore2012}. The intricate of geometrical frustration and long-range electron-mediated interactions leads to a rich phase diagram with complex non-collinear magnetic orders controlled by the strength of Hund's coupling $J$ and the electron filling fraction. 

Fully connected NNs are used as learning models in all cases discussed below. The NN was implemented using PyTorch~\cite{Paszke2019} and consists of eight hidden layers with neuron sizes structured as $2048\times1024\times512\times256\times128\times64\times64\times64$ neurons, respectively. The input layer of the model was determined by the number of feature variables $\{G_\ell\}$, which is set at 1806 determined by the number of feature variables within the neighborhood with $r_c = 6a$ and $l_c = 2a$, where $a$ is the lattice constant.  The output layer of the model consisted of a single neuron whose value corresponds to the predicted local energy $\epsilon_i$. Rectified linear unit (ReLU)~\cite{Nair2010} activation function was employed between layers of the NN model. Given that the torque $\textbf{T}_i=\textbf{S}_i\times\textbf{H}_i$ is the key component for deriving the LLG dynamics, the mean square error (MSE) of both the total energy and the local effective fields are used as the loss function
\begin{equation}
	\mathcal{L}(\bm \theta) = \sum^N_{i=1} \bigl| \mathbf T_i - \hat{\mathbf T}_i \bigr|^2
	+ \eta_E \Bigl| E - \sum_i^N \hat{\epsilon}_i \Bigr|^2. \,\,
	\label{eq:loss_func}
\end{equation}
Here quantities with a hat denote predictions from the ML models, and are dependent on the trainable parameters $\bm \theta$. The optimization of these parameters is carried out using the Adam stochastic optimization algorithm~\cite{Kingma2014} with a learning rate that undergoes exponential decay as the number of training iterations increases, starting at 0.1 and ending at 0.0001. To train the neural network, a dataset consisting of 40,000 snapshots derived from exact simulations is utilized, and training is conducted over 200 epochs. In order to mitigate overfitting, a 5-fold cross-validation strategy is employed~\cite{Stone1974}.

The trained ML models were then integrated with the LLG simulations and benchmarked with simulations with local fields computed using either the KPM or ED. A semi-implicit scheme was used to integrate the stochastic LLG equation in the following simulations~\cite{Mentink10}.   Since the length of classical spins is set to $|\mathbf S_i| =1$,  the gyromagnetic ratio does not have the conventional units. Instead, as the effective field has the unit of energy, $\gamma$ has the same unit as the inverse Planck's constant, $1/\hbar$. Practically, both $\gamma$ and the nearest-neighbor hopping $t_{\rm nn}$ are set to 1 in our simulations which means that the simulation time in the following is measured in units of $\tau_0 = (\gamma t_{\rm nn})^{-1}$. In the following, a time-step $\Delta t = 0.025 \tau_0$ and a dimensionless dissipation coefficient $\alpha = 0.075$ are used in all simulations.  

\begin{figure*}[t]
\includegraphics[width=2.0\columnwidth]{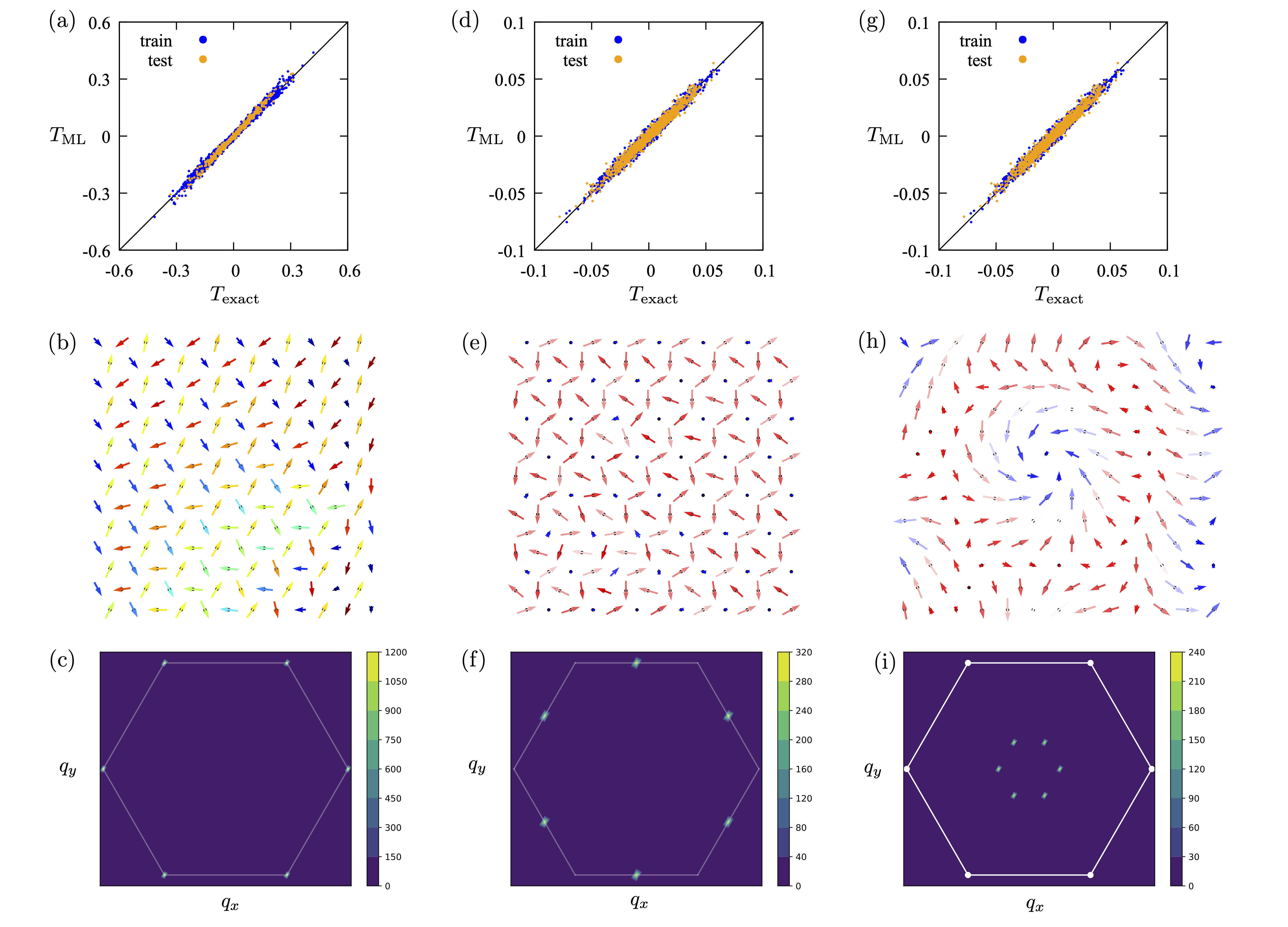}
\caption{Top row: Comparison between the component-wise ML-predicted torques $\mathbf T_{\rm ML} = \mathbf S \times \mathbf H_{\rm ML}$ and the ground truth $\mathbf T_{\rm exact}$ for s-d models that exhibit (a) the coplanar 120$^\circ$ order, (d) the non-coplanar tetrahedral order, and (g) a skyrmion crystal phase in the ground state. The blue and orange data points correspond to training and test datasets, respectively. The three panels (b), (e), (h) in the middle display the corresponding snapshots of spin configurations obtained from ML-based LLG simulations for these three cases. The bottom panels (c), (f), (i) show the ensemble-averaged spin structure factor $\mathcal{S}(\mathbf q)$ of these three phases from ML-LLG simulations. }
\label{fig:ml-summary}  
\end{figure*}

\subsection{Coplanar 120$^\circ$ order }

We first consider the s-d model with only nearest-neighbor hopping $t_{\rm nn}$ and a Hund's coupling $J = 6 \, t_{\rm nn}$ at half filling, which also serves as a benchmark of our approach. To understand the magnetic ground states, we first note that a half-filled s-d model in the strong coupling limit $J \gg t_{\rm nn}$ can be mapped to a classical antiferromagnetic Heisenberg model, regardless of the underlying lattice~\cite{Chern18}. To see this, we note that any ground state in the $t_{\rm nn} = 0$ limit contains exactly one electron in each site with its spin aligned with the local moment~$\mathbf S_i$. Similar to the large $U$ limit of a half-filled Hubbard model~\cite{Fazekas99}, the ground-state manifold is massively degenerate because each~$\mathbf S_i$ can point in an arbitrary direction. The effects of a small electron hopping can be treated using the degenerate perturbation theory, which at the leading second order leads to an effective energy between a spin pair $E_{ij} = -2 (t^{\rm eff}_{ij})^2 / J$, where the effective hopping is $t^{\rm eff}_{ij} = t_{\rm nn} \sin(\theta_{ij} / 2)$ and $\theta_{ij}$ is the angle between the two nearest-neighbor local moments $\mathbf S_i$ and~$\mathbf S_j$. Straightforward calculations lead to a Heisenberg Hamiltonian $E = J_{\rm eff} \sum_{\langle ij \rangle} \mathbf S_i \cdot \mathbf S_j + \mbox{const.}$, with an effective exchange interaction $J_{\rm eff} = {4 t_{\rm nn}^2}/{J} > 0$. For triangular lattice, the nearest-neighbor antiferromagnetic exchange interactions cannot be satisfied simultaneously on every triangle units, a classical case of geometrical frustration~\cite{lacroix2011}. The resultant classical ground state is the well-known 120$^\circ$ order with a tripled unit cell: 
\begin{eqnarray}
	\mathbf S_i =\cos(\mathbf K \cdot \mathbf r_i) \, \hat{\mathbf n}_1 +  \sin(\mathbf K \cdot \mathbf r_i) \, \hat{\mathbf n}_2, 
\end{eqnarray}
where $\hat{\mathbf n}_1$ and $\hat{\mathbf n}_2$ are two orthogonal unit vectors, and $\mathbf K = (4\pi /3 a, 0)$ is the ordering wave vector. 

Applying the ML force-field approach to this case, we trained our NN model using a total of 40000 spin configurations, including the several stages of the phase ordering process, from 40 independent ED-LLG simulations. The training dataset was obtained from KPM-based LLG simulations on a relatively small $48\times 48$ lattice. Fig.~\ref{fig:ml-summary}(a) shows the comparison of the ML-predicted componentwise torques versus the ground truth for both the training and test datasets. A mean squared error (MSE) of $8.79\times10^{-8}$ was achieved from our analysis, indicating excellent accuracy of the trained ML model. Moreover, the close proximity of the errors between the test and training datasets indicate that the NN models exhibit minimal overfitting. We next utilize LLG method with ML predicted local fields to simulate thermal quench scenario on a $150\times 150$ system.  The system initially prepared in a state of random spins is coupled to a thermal bath, which is suddenly quenched to a low temperature. Fig.~\ref{fig:ml-summary}(b) shows a snapshot of spins near the end of phase ordering. A clear three-sublattice structure can be observed with spins on every triangle unit assuming the nearly 120$^\circ$ configuration. We further compute the spin structure factor, which is given by the fourier transform of the spin field
\begin{eqnarray}
	\mathcal{S}(\mathbf q, t) = \Bigl| \frac{1}{N} \sum_i \mathbf S_i(t) e^{i \mathbf q \cdot \mathbf r_i} \Bigr|^2,
\end{eqnarray}
As shown in Fig.~\ref{fig:ml-summary}(c), the structure factor exhibits clear peaks at the ordering wave vector $\mathbf K$, which corresponds to the six corners of the Brillouin zone (BZ).

\subsection{Noncoplanar tetrahedral order}

Next we consider the noncoplanar tetrahedral order with a quadrupled unit cell, which are the ground state of the triangular-lattice s-d model at filling fraction $f = 3/4$ and $f \sim 1/4$~\cite{Martin2008,Akagi10,Kato2010,Azhar17,Chern2012}.  This magnetic order has a quadrupled unit cell with local spins pointing toward the corners of a regular tetrahedron, as shown in Fig.~\ref{fig:ml-summary}(e). In general, a magnetic order with a quadrupled unit cell on the triangular lattice is described by a triple-$\mathbf Q$ magnetic structure:
\begin{eqnarray}
	\label{eq:triple-Q}
	\mathbf S_i = \bm \Delta_1 e^{i \mathbf M_1 \cdot \mathbf r_i} + \bm \Delta_2 e^{i \mathbf M_2 \cdot \mathbf r_i} + 
	\bm \Delta_3 e^{i \mathbf M_3 \cdot \mathbf r_i},
\end{eqnarray}
where the three commensurate ordering wave vectors are $\mathbf M_1 = (2\pi/a, 0)$, and $\mathbf M_{2, 3} = (-\pi/a, \pm \sqrt{3} \pi/a)$, corresponding to the mid-points of the edge of the hexagonal BZ, and $\bm\Delta_\eta$ is the vector order parameters associated with wave vector $\mathbf M_\eta$. The highly symmetric tetrahedral spin order corresponds to the special case when the three vector order parameters are orthogonal to each other and have the same amplitude, i.e. $\bm \Delta_\eta = \Delta \,\hat{\mathbf e}_{\eta}$, where $\Delta$ is a real number characterizing the amplitude of spin order and $\hat{\mathbf e}_\eta$ are three mutually orthogonal unit vectors. It is straightforward to check that the four spins in the quadrupled unit cell point to the corners of a regular tetrahedron.

The stabilization mechanism for the tetrahedral order, however, is very different for the two filling fractions. For the case of 3/4-filling, it is stabilized by a perfect Fermi surface nesting~\cite{Martin2008,Chern2012}, similar to the scenario of N\'eel order in square-lattice Hubbard or s-d models at half filling. 
The emergence of the tetrahedral spin order at $n \sim 1/4$, on the other hand, is unexpected from the Fermi surface nesting scenario. The Fermi surface in the vicinity of quarter filling has an overall circular shape, with no substantial nesting tendency. The magnetic susceptibility $\chi(\mathbf q)$, computed from the second-order perturbation expansion in $J/t$, does show a weak maximum at the three commensurate wave vectors $\mathbf M_\eta$ for filling fraction $n \sim 1/4$. Yet, the single, double, and triple-$\mathbf Q$ orderings remain degenerate in energy at this order. This degeneracy is lifted at the fourth order, where a positive biquadratic spin interaction $B(\mathbf S_i \cdot \mathbf S_j)^2$ with a positive coefficient $B > 0$ clearly favors the tetrahedral triple-$\mathbf Q$ order~\cite{Akagi2012}. 

Additionally, the three ordering wave vectors $\mathbf Q_\eta$ link specific points on the Fermi surface at a filling fraction of $f \sim 1/4$, where the tangents to the Fermi surface at these points are parallel to each other. This unique geometry induces a singularity similar to the $2 k_F$ Kohn anomaly seen in an isotropic electron gas. Crucially, this singularity causes a significant enhancement of the positive biquadratic interaction $B$, which in turn stabilizes the noncoplanar tetrahedral spin order~\cite{Akagi2012}.

\begin{figure}[t]
\includegraphics[width=0.99\columnwidth]{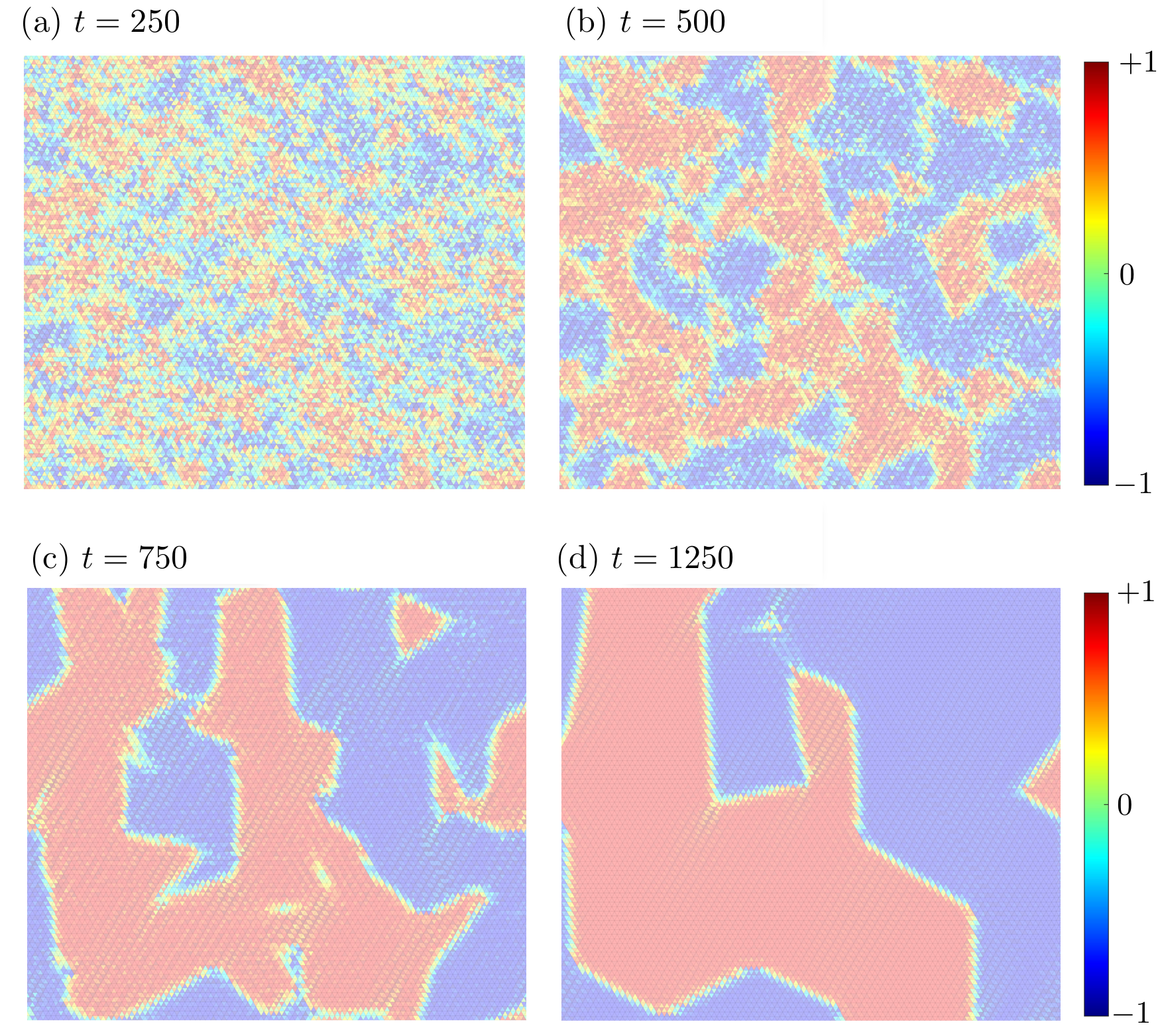}
\caption{Snapshots of local scalar chirality $\chi_{ijk} = \mathbf S_i \cdot \mathbf S_j \times \mathbf S_k$, where $(ijk)$ denotes lattice sites on an elementary triangular unit, at various time steps after a thermal quench of the triangular s-d model at filling fraction close to 1/4.}
\label{fig:chiral-coarsening}  
\end{figure}

\begin{figure*}[t]
\includegraphics[width=2.0\columnwidth]{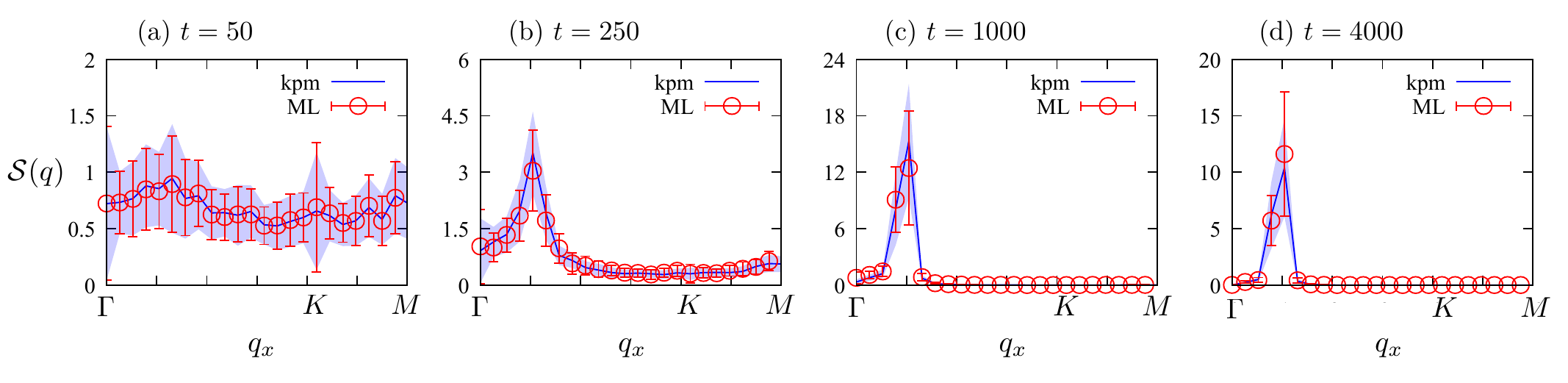}
\caption{Comparison of time-dependent structure factors $\mathcal{S}(\mathbf q)$ obtained from KPM- and ML-based LLG simulations on a $48\times 48$ lattice. The structure factors at different times after a thermal quench are plotted along high symmetry directions $\Gamma(0,0)\rightarrow K(\frac{4\pi}{3},0)\rightarrow M(2\pi,0)$. The blue lines show values of $\mathcal{S}(\mathbf q)$ averaged over 50 independent KPM-LLG simulations, with the shaded area indicating the standard deviations.  The red dots represent the average $\mathcal{S}(\mathbf q)$ from ML-LLG simulations with the corresponding standard deviation indicated by the error bars. Note the difference in $y$ scales at different times.}
\label{fig:dyn-benchmark}  
\end{figure*}

Here we built a ML force-field model to learn the dynamics of the s-d system with $J/t_{\rm nn} = 3$. The training dataset was obtained using KPM within the grand-canonical ensemble framework with the electron fermi energy set at $\mu = -3.2 t_{\rm nn}$ which corresponds to a filling fraction in the vicinity of 1/4. The benchmark of the force prediction shown in Fig.~\ref{fig:ml-summary}(d) indicates good agreement with a MSE of $7.80\times 10^{-7}$ and no evidence of overfitting. Further dynamical benchmark can be found in Ref.~\cite{Fan24}.  A snapshot of spin configuration at late stage of the phase ordering clearly shows a four-sublattice structure with a small fluctuations, consistent with the tetrahedral order; see Fig.~\ref{fig:ml-summary}(e). The triple-$\mathbf Q$ structure of the tetrahedral order is also confirmed by the ensemble-averaged structure factor, shown in Fig.~\ref{fig:ml-summary}(f), which shows six sharp peaks at the six mid-points of the BZ edges; these $M$ points correspond to the $\pm \mathbf M_\eta$ ordering wave vectors.

Before concluding this section, we note that the noncoplanar spins in the tetrahedral order also breaks the $Z_2$ chiral symmetry, which can be characterized by a nonzero scalar spin chirality $\chi_{ijk} =  \mathbf S_i \cdot \mathbf  S_j \times \mathbf S_k $ for three spins on an elementary triangular unit.  Remarkably, the chiral ordered state exhibits a quantized Hall conductivity of $\pm e^2/h$, with the sign determined by that of the scalar chirality~\cite{Martin2008,Chern2012}. It is worth noting that, while the long-range tetrahedral spin order is unstable against thermal fluctuations in 2D, the chirality order associated with a broken $Z_2$ symmetry persists at finite temperatures. On symmetry grounds, the chiral order can be described by an Ising-type order parameter. Indeed, as shown in Fig.~\ref{fig:chiral-coarsening}, the phase ordering dynamics of the 1/4-filled s-d model is characterized by the coarsening of Ising-type domains with well-defined chirality. Interestingly, our ML-enabled large-scale LLG simulations uncover a linear growth of the chiral domains~\cite{Fan24}, in stark contrast to the expected Allen-Cahn domain growth law for a non-conserved Ising order parameter field~\cite{Bray1994,Puri2009,Onuki2002}. The linear growth of the chiral domains is attributed to the orientational anisotropy of domain boundaries.

\subsection{Triple-$\mathbf Q$ Skyrmion lattice}

As a last example, we apply our ML force-field framework to study the dynamics of a skyrmion crystal phase of the s-d model. As discussed above, skyrmion crystals can be stabilized by the quasi-nesting mechanism in s-d model through the engineering of the electron Fermi surface. Here we consider a tight-binding model for $\hat{\mathcal{H}}_e$ with a third-nearest-neighbor hopping $t_3 = -0.85 t_{\rm nn}$ in addition to the nearest-neighbor one, and a Hund's coupling of $J = t_{\rm nn}$ studied in Ref.~\cite{Ozawa2017}. The electron chemical potential is set at $\mu = -3.5 t_{\rm nn}$, corresponding to a filling fraction slightly smaller than 1/5. An additional Zeeman coupling to classical spins 
\begin{eqnarray}
	E_{\rm Zeeman} = -  \sum \mathbf H_{\rm ext} \cdot \mathbf S_i, 
\end{eqnarray}
with a small magnetic field $|\mathbf H_{\rm ext}| = 0.005 t_{\rm nn}$ is added to the s-d Hamiltonian in order to stabilize the skyrmion phase. We note that this Zeeman term, which does not involve electron degrees of freedom, is not included in the training of the ML models. In practical simulations, a constant external field $\mathbf H_{\rm ext}$ along the $z$ direction is added to the driving forces of LLG equation.

The geometry of the Fermi surface of this s-d model allows significant segments to be connected by three wave vectors $\mathbf Q_1 = (\pi / 3a, 0)$ and $\mathbf Q_{2, 3} = \mathcal{R}_{\pm 2\pi/3} \cdot \mathbf Q_1$, related to each other by $\pm 120^\circ$ rotation which are part of the $D_6$ point group. Here $a$ is the lattice constant of the underlying triangular lattice. This means that maximum energy gain through electron-spin coupling is realized by spin helical orders with one of the above three wave vectors. The electron energy is further lowered by the simultaneous ordering of all three wave vectors. The emergent skyrmion crystal of this triple-$\mathbf Q$ magnetic order can be approximated by the following~\cite{Ozawa2017,Hayami2021}
\begin{eqnarray}
	\label{eq:skl}
	& & \mathbf S_i \sim  \left( \cos\mathcal{Q}_{1i} - \frac{1}{2} \cos\mathcal{Q}_{2i} - \frac{1}{2} \mathcal{Q}_{3i} \right) \hat{\mathbf e}_1 \nonumber \\
	& & \qquad + \left( \frac{\sqrt{3}}{2} \cos\mathcal{Q}_{2i} - \frac{\sqrt{3}}{2} \cos\mathcal{Q}_{3i} \right) \hat{\mathbf e}_2  \\
	& & \qquad +  \left[ A \left(\sin\mathcal{Q}'_{1i} + \sin\mathcal{Q}'_{2i} + \sin\mathcal{Q}'_{3i} \right) + M \right] \hat{\mathbf e}_3, \nonumber
\end{eqnarray}
where $\hat{\mathbf e}_{1, 2, 3}$ are three orthogonal unit vectors, $\mathcal{Q}_{\eta i} = \mathbf Q_\eta \cdot \mathbf r_i$, and $\mathcal{Q}'_{\eta, i} = \mathcal{Q}_{\eta, i} + \phi$ are phase factors of the three helical orders, $\phi$, $A$, and $M$ are fitting parameters. The ordering wave vectors here are designed to be commensurate with the underlying triangular lattice. As a result, the skyrmion crystals correspond to a magnetic order with a enlarged unit cell with 12 spins. 

The NN model for the skyrmion phase is again trained by dataset of 40,000 spin configurations, including different stages of the phase ordering as well as perturbed skyrmion crystal state, from 40 independent simulations. Benchmark of the force prediction, shown in Fig.~\ref{fig:ml-summary}(g), indicates overall agreement with the exact results, with a fairly small MSE of $9.82\times10^{-7}$ and no signs of overfitting. LLG simulations based on ML-predicted local fields also showed the emergence of swirling spin texture of which the spins wrap around a unit sphere, as shown in Fig.~\ref{fig:ml-summary}(h), characteristic of a skyrmion. 


We next performed a dynamical benchmark of the trained ML model. To this end, we compared the evolutions of the spin structure factor at different times after a thermal quench using LLG simulations with local fields obtained by KPM and ML predictions. For both methods, the structure factors were computed from averaging over 50 independent simulations on a relatively smaller $48\times 48$ lattice. Quantitative agreements were obtained for the resultant time-dependent structure factors, which are shown in Fig.~\ref{fig:dyn-benchmark} along high symmetry directions of the BZ: $\Gamma(0,0)\rightarrow K(\frac{4\pi}{3},0)\rightarrow M(2\pi,0)$. In the initial stage shown in Fig.~\ref{fig:dyn-benchmark}(a), strong fluctuations were observed at the $K$ points, indicating a $\sqrt{3}\times \sqrt{3}$ correlation related to the 120$^\circ$ order. As the system relaxed toward equilibrium, strong peaks quickly develop at the wave vectors $\mathbf Q_1 = (\pi / 3, 0)$, highlighting the emergence of the triple-$\mathbf Q$ skyrmion structures.

\section{Arrested ordering kinetics of Skyrmion phases}

\begin{figure}[t]
\includegraphics[width=0.9\columnwidth]{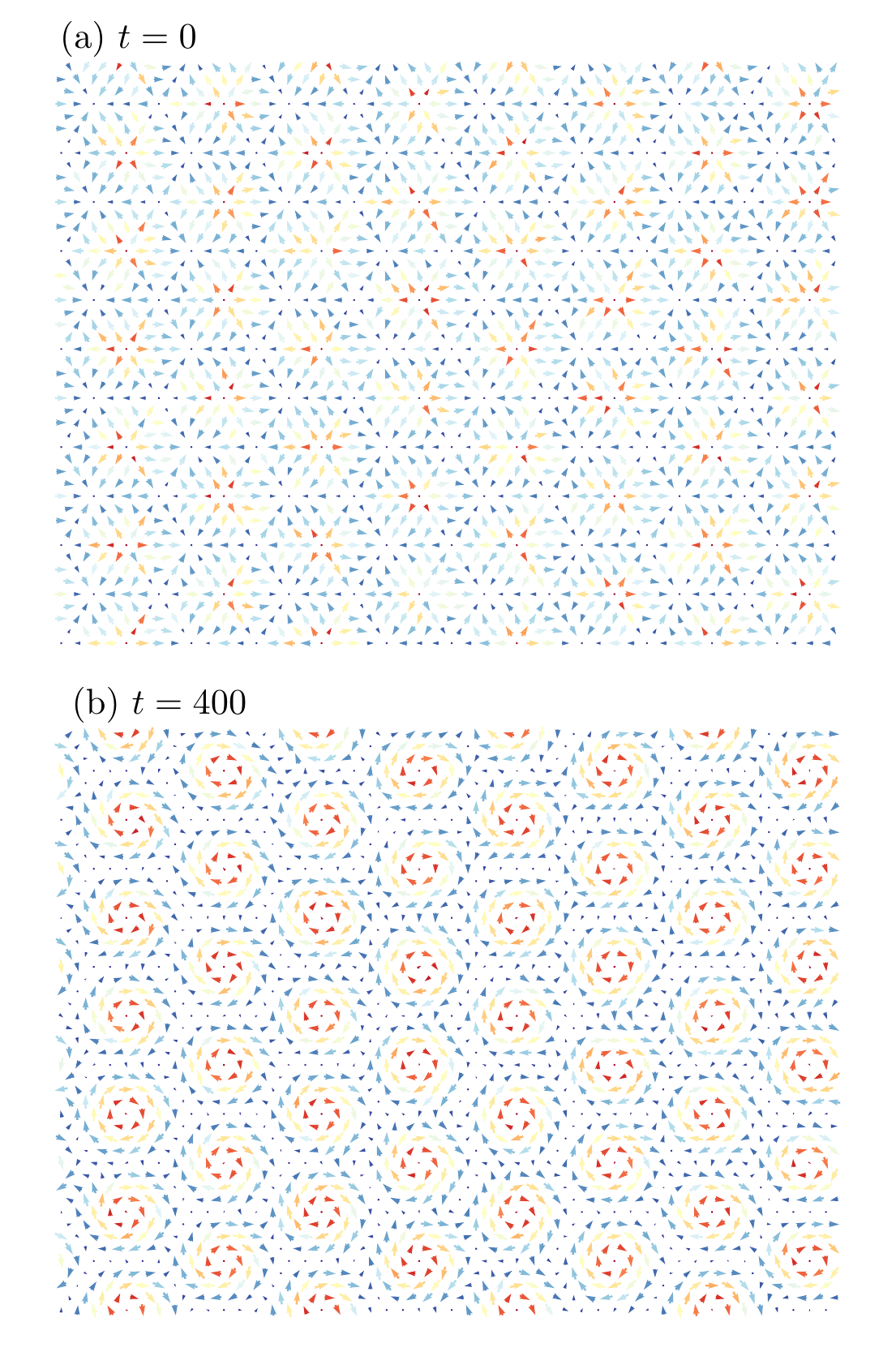}
\caption{ML-based LLG simulation on a $48\times 48$ lattice showing the restoration of a perturbed skyrmion crystal. The spins are initialized using Eq.~(\ref{eq:skl}) with additional site-dependent random phases and amplitudes of $S^z$. }
\label{fig:skx-benchmark}  
\end{figure}

\begin{figure*}[t]
\includegraphics[width=2.0\columnwidth]{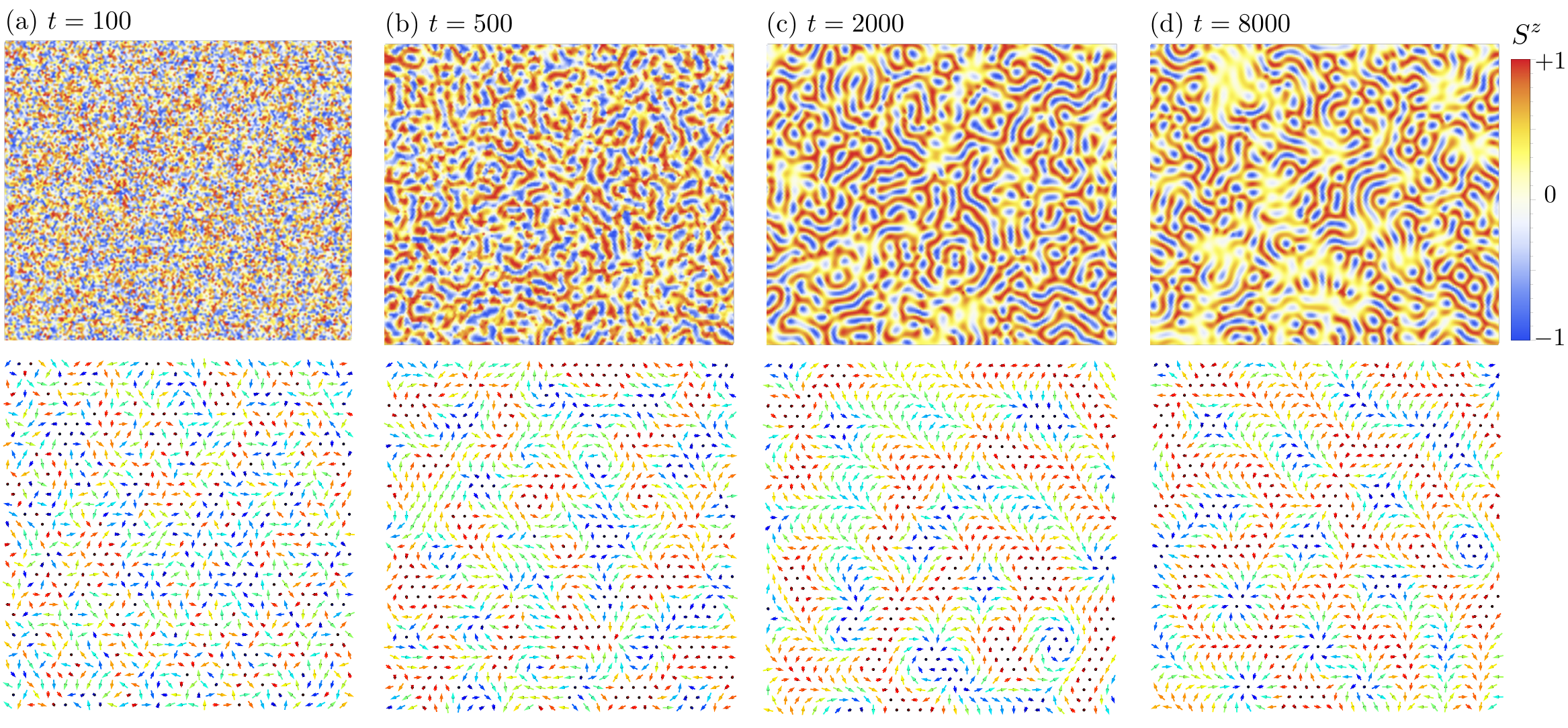}
\caption{Snapshots of spin configurations obtained from ML-LLG simulations of thermal quench dynamics. The system size is $150\times 150$. Top panels show distribution of the $S^z$ component at different times after a thermal quench, while the corresponding $(S^x, S^y)$ components spins in a $32\times 32$ block at the top right corner of the system are shown in the bottom vector plots, with the $S^z$ amplitude indicated by the color of arrows.  }
\label{fig:thermal-quench}  
\end{figure*}

In this section we apply the trained ML model to investigate the phase ordering dynamics of skyrmion crystals. As a further benchmark, we first show that the skyrmion crystal is indeed a stable minimum energy state of the ML force-field model. To this end, we first create a perfect array of skyrmions described by the ansatz Eq.~(\ref{eq:skl}) and perturb it by introducing small fluctuations to the parameters $\phi_i$, $A_i$ and $m_i$. The resultant perturbed skyrmion crystal, shown in Fig.~\ref{fig:skx-benchmark}(a), serves as the initial state for the LLG simulations. By setting the Langevin thermostat at $T = 10^{-4}$ and a dimensionless dissipation $\lambda = 0.2$, our LLG simulation with local fields predicted from the ML model found that a nearly perfect skyrmion crystal is restored and stabilized over a long period of simulation time; see Fig.~\ref{fig:skx-benchmark}(b). The spin structure factor $\mathcal{S}(\mathbf q)$ obtained from the relaxed skyrmion lattice is shown in Fig.~\ref{fig:ml-summary}(i). The six sharp peaks at the designed order wave vectors again indicate the underlying tripe-$\mathbf Q$ structure of the skyrmion lattice. 

Next we simulate the standard thermal quench scenario in which a system with initial random spins is quenched into the low temperature skyrmion phase through a Langevin thermostat. Fig.~\ref{fig:thermal-quench} shows snapshots of spin configurations obtained from the ML-LLG simulation at different times after a quench to a low temperature $T = 10^{-4}$. The top panels show the $S^z$ component of spins, while the vector plots in the bottom row depicts the in-plan $(S^x, S^y)$ components.A skyrmion here is characterized by isolated circular blue region of negative $S^z$ accompanied by a circular vortex structure in the $S^x$ and $S^y$ components.  Interestingly, although several skyrmion crystallites, i.e. small patches of ordered skyrmion arrays, are stabilized during the equilibration process, no large coherent crystals emerge even at the late stage of the relaxation.

Due to the 2D nature of skyrmion lattices, their crystallization process is intimately related to the well-known Kosterlitz-Thouless-Halperin-Nelson-Young (KTHNY) theory for the melting of 2D crystals~\cite{Kosterlitz1972,Kosterlitz1973,Halperin1978,Nelson1979,Young1979}.  In line with the modern theory of phase ordering~\cite{Bray1994}, topological defects associated with translational and rotational symmetry breaking called dislocations and disclinations, respectively, play a crucial role in the melting of 2D crystals. The KTHNY theory posits a two-stage scenario with a partially ordered phase in between the solid and isotropic fluid phases. The intermediate phase, called the hexatic phase for systems with six-fold rotation symmetry, is characterized by a quasi-long-range rotational order and a short-range positional order. The crystallization dynamics is thus expected to be dominated by the annihilation of these two types of topological defects.


Based on the above framework, a plausible scenario for the phase ordering of skyrmion lattice is as follows. The first stage is the formation of individual skyrmions as quasi-particles, which requires orientational arrangements of spins in a local region. This part has no analogy in crystallizations of particles (atoms or molecules) since skyrmions are emergent particle-like objects. The formation of large skyrmion crystals is expected to follow a similar two-stage process: the breaking of rotational symmetry via the elimination of disclination defects, followed by the breaking of the continuous translation symmetry assisted by the pair annihilation of dislocations. However, this scenario does not appear to be supported by our ML-LLG simulations, summarized in Fig.~\ref{fig:thermal-quench}. Even in the later stages of the relaxation process, no well-defined disclination or dislocation defects are observed. 

Indeed, our simulations found that several bimerons persist even in the very late stages of the relaxation. A compact bimeron consists of two half-disk-like domains and a rectangular stripe domain, with spins oriented upward at the boundary and downward toward the inside of the bimeron~\cite{Ezawa11}. As each of the half-disk carries one half of the skyrmion number, a bimeron can be viewed as a quasi-fractionalized elongated skyrmion. This indicates that even the initial stage of the particle-based scenario -- specifically, the proper formation of skyrmion quasi-particles -- is incomplete.

\begin{figure*}[t]
\includegraphics[width=2.0\columnwidth]{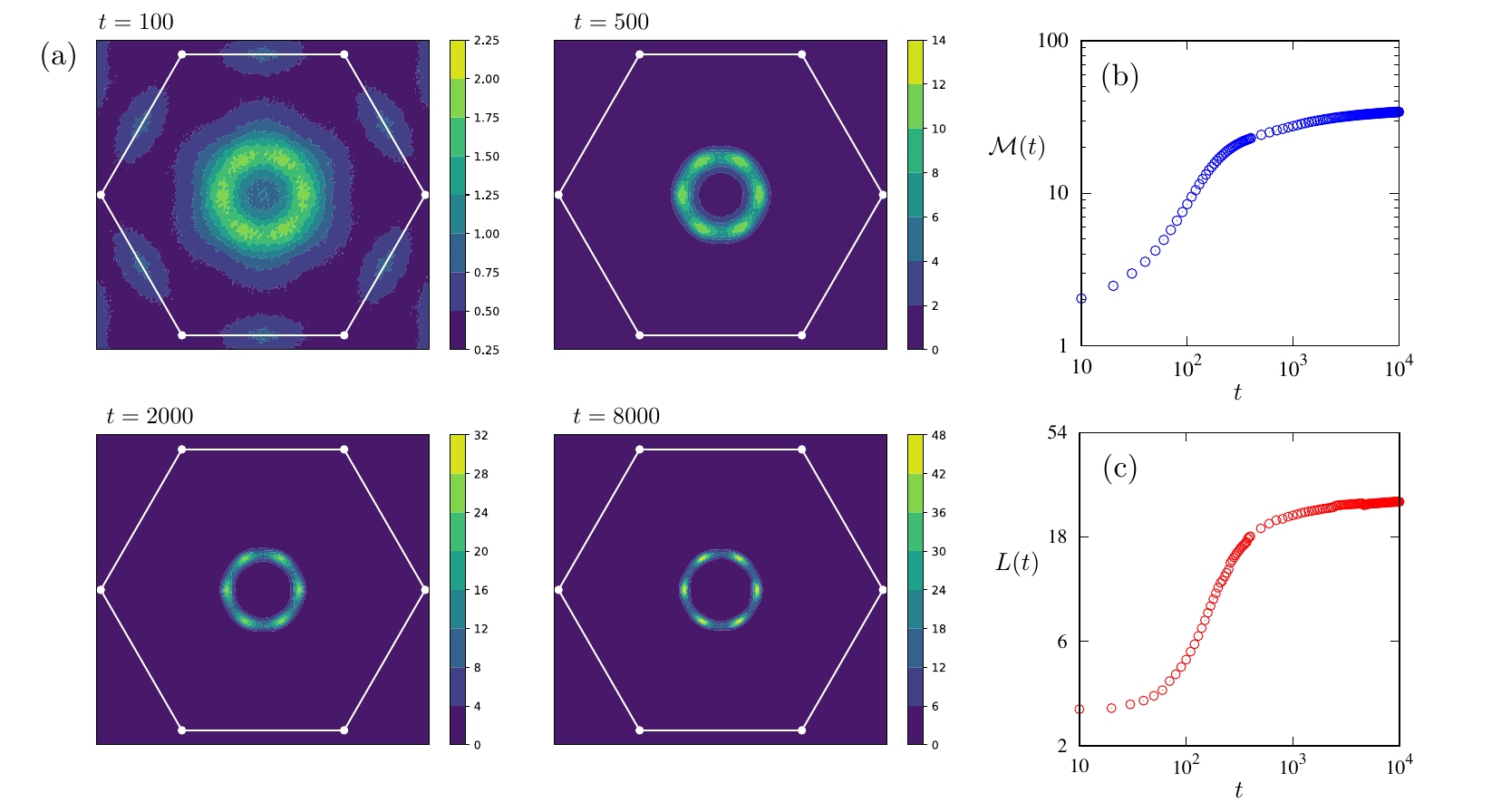}
\caption{Left four panels (a): The structure factor density plots of the ML-LLG  $150\times150$ lattice size simulations in the first Brillouin zone from $t=50$ to 4000. (b) The skyrmion structure factor peak order $\mathcal{M}$ in the different timestep from $t=10$ to $10^4$. (c) The characteristic length of the ML-LLG simulations from $t=10$ to $10^4$.}
\label{fig:ordering-dyn}  
\end{figure*}

Instead, the numerous  stripe-like structures that dominate the quenched states indicate the emergence of stable single-$\mathbf Q$ helical order. As the skyrmion crystal corresponds to a triple-$\mathbf Q$ order, which can be viewed as the coexistence of three helical orders running along the three principal directions of the triangular lattice. The phase ordering dynamics is thus better characterized by the competition between the single-$\mathbf Q$ and triple-$\mathbf Q$ ordering. Indeed, using high precision KPM, we found that the single, double, and triple-$\mathbf Q$ states are nearly degenerate in energy, with a slightly lowered energy for the triple-$\mathbf Q$ skyrmion crystal. To characterize this process in momentum space, we plot the ensemble-averaged structure factor $\mathcal{S}(\mathbf q)$ at different times after the quench in Fig.~\ref{fig:ordering-dyn}. Instead of six sharp peaks at the ordering wave vectors $\pm \mathbf Q_\eta$, the structure factor exhibits a prominent ring-like structure with its radius equal to the magnitude of the ordering wave vectors. 

The ring-like feature is also well-known in the so-called labyrinthine structures in thin-film magnets. These maze-like structures, characterized by alternating up and down regions of magnetization, arranged in irregular, curving pathways, are clearly visible in our simulations. The prevalence of these stripes is partly attributed to their representation of local energy minima within the system, and partly because of their favored configurational entropy compared with the skyrmion crystals.

Finally, the phase ordering is clearly arrested as the system is frozen in the disordered configurations. To quantify the freezing behavior, we introduce an order parameter corresponding to the overall triple-$\mathbf Q$ ordering
\begin{eqnarray}
	\mathcal{M}(t) = \sum_{\eta = 1,2,3}\sum_{|\mathbf q \pm  \mathbf Q_\eta| \le \delta q} \mathcal{S}(\mathbf q, t)
\end{eqnarray}
This parameter essentially measures the combined momenta within a radius $\delta q$, which is chosen to be 0.1 in our calculation, around the six ordering wave vectors.  As shown in Fig.~\ref{fig:ordering-dyn}(b), the order parameter $\mathcal{M}$ grows quickly at first, then gradually levels off at late stages of the equilibration. This arrested growth of the spin order at favored $\mathbf Q_\eta$ indicates that a significant fraction of the momentum distribution is frozen over the ring. 
In addition to the overall triple-$\mathbf Q$ order, one can also characterize the size of the skyrmion crystallites. This correlation length of the triple-$\mathbf Q$ structure can be estimated from the inverse width of the diffusive peaks at the six ordering wave vectors:
\begin{eqnarray}
	L^{-1}(t) = \frac{ \sum_{\eta = 1, 2, 3} \sum_{ \mathbf q} |\mathbf q \pm \mathbf Q_\eta| \mathcal{S}(\mathbf q, t) }{ \sum_{\mathbf q} \mathcal{S}(\mathbf q, t) }.
\end{eqnarray}
The time dependence of this length scale shown in Fig.~\ref{fig:ordering-dyn}(c) indicates a similar arrested growth of skyrmion crystallites. 

An emerging picture from our large-scale ML-LLG simulations is the following. Rather than first forming skyrmion quasi-particles that crystallize in two steps through the annihilation of disclination and dislocation defects, small patches of skyrmion arrays emerge directly through triple-$\mathbf{Q}$ ordering, achieved by locking the order parameters of the three ordering wave vectors. However, the incompatibility due to incoherent growths of these skyrmion crystallites lead to formation of stripe structures, with isolated skyrmions and bimerons, that interpolate between areas of skyrmion lattice. The stability of these intervening stripes or single-$\mathbf Q$ helical spins induces an arrested domain growth.

\section{Discussion}

In summary, we present a comprehensive review of the recent advancements in machine-learning (ML) force-field frameworks for simulating Landau-Lifshitz-Gilbert (LLG) dynamics in itinerant electron magnets. By leveraging the locality principle, which posits that the electronic forces acting on the spins are dependent on their local environment, we developed a deep learning neural network capable of accurately encoding the complex and nonlinear relationship between local spin configurations and the resulting effective magnetic field.  A crucial component of the proposed ML force field models is the magnetic descriptors, which refers to a symmetry-invariant representations of a magnetic environment. These representations must satisfy critical properties: they should be differentiable with respect to spin rotations, and invariant to both lattice point-group symmetry and internal spin rotation symmetry.  We discuss the background theory and practical implementations of magnetic descriptors. focusing on the group-theoretical approaches. In particular, we introduce an efficient implementation built around the concept of reference irreducible representations, which is an adaptation of the group-theoretical power-spectrum and bispectrum methods. 

The proposed ML framework is demonstrated through its application to the s-d models, a widely-used system in spintronics research. We show that LLG simulations with local fields predicted by the trained ML models accurately reproduce key non-collinear spin structures such as 120$^\circ$ order, tetrahedral configurations, and skyrmion crystal arrangements within the triangular-lattice s-d models. Furthermore, our large-scale ML-LLG simulations uncover intriguing arrested phase ordering of skyrmion crystals and the emergence of a glassy skyrmion state. Our findings underscore the potential of the ML force-field approach for modeling the dynamics of complex spin configurations in itinerant electron magnets, highlighting its ability to capture intricate spin interactions and provide insight into the behavior of these systems under various conditions. 

In itinerant systems with explicit spin-orbit interactions, such as the Rashba-type tight-binding model, the SO(3) spin-rotation symmetry is intrinsically coupled with the real-space lattice point-group symmetry. Nonetheless, the ML force-field framework presented here remains applicable to these anisotropic systems. The primary modification lies in developing a magnetic descriptor that accommodates the combined spin-lattice symmetry, as opposed to treating the spin-rotation and lattice point-group symmetries independently. Nevertheless, the group-theoretical method and the concept of reference IR introduced in this work can still be effectively applied to handle the combined spin-lattice symmetry group.

The ML force-field models discussed in this work are essentially the magnetic generalization of the Behler-Parrinello architecture. Two notable features of the BP scheme are the explicit utilization of locality principle and the introduction of a local energy. The latter also facilitates the incorporation of symmetry constraints into the ML force-field models. An alternative approach is based on convolutional neural networks (CNNs), which are a class of neural networks defined by their local connectivity through finite-sized convolution kernels. Notably, the convolution operation with a finite kernel also inherently embeds the locality principle within the ML architecture, providing a scalable realization of ML force-field models. A CNN-based force-field model for spin dynamics was recently demonstrated~\cite{cheng23b,Miyazaki23}. However,  integrating symmetry constraints into CNN-based approaches is a challenging task. One approach is the conventional data augmentation techniques to approximately account for both the spin-rotation and the lattice symmetries~\cite{cheng23b}. 

The recent advent of equivariant neural networks~\cite{weiler18,batzner22,gong23} offers a tantalizing alternative to descriptor-based ML models. An equivariant NN is designed in such a way that neurons at each layer transform according to well-defined rules under a given symmetry group. In the context of spin dynamics, it has been demonstrated that the SO(3) spin rotation symmetry can be nicely incorporated into a ML force-field model via an equivariant CNN~\cite{Miyazaki23}. Yet, it remains unclear how to consistently treat both spin rotation symmetry and lattice point-group symmetry within an equivariant NN framework, which will be left for future studies.

\bibliography{ref.bib}

\end{document}